\documentclass[a4paper,11pt]{article}
\usepackage{jcappub} 
\usepackage{physics}
\usepackage{multirow}
\usepackage{aas_macros}
\usepackage{newtxtext,newtxmath}
\usepackage{multirow}
\usepackage[T1]{fontenc}
\usepackage{ae,aecompl}
\usepackage{graphicx}	%
\usepackage{amsmath}	%
\usepackage{amssymb}	%
\usepackage{multicol}	%
\usepackage{bm}		%
\usepackage{pdflscape}	%
\usepackage{hyperref}
\usepackage{lineno} %
\usepackage{lipsum}
\usepackage{mathtools}

\usepackage{tikz}
\usetikzlibrary{shapes.geometric, arrows, shapes.arrows, calc}

\tikzstyle{input} = [ellipse, 
minimum width=2cm, 
minimum height=1cm, 
text centered, 
text width=2cm, 
draw=yellow!80, 
fill=yellow!30]

\tikzstyle{process} = [rectangle, 
minimum width=1cm, 
minimum height=1cm, 
text centered, 
text width=3cm, 
draw=blue!80, 
fill=blue!30]

\tikzstyle{model} = [circle, 
minimum width=2cm, 
minimum height=1cm, 
text centered, 
text width=2cm, 
draw=green!80, 
fill=green!30]

\tikzstyle{arrow} = [thick,->,>=stealth]

\usepackage{graphicx}	%
\usepackage{amsmath}	%
\usepackage{amssymb}	%

\arxivnumber{1234.56789} 
\title{Deciphering Baryonic Feedback from ACT tSZ Galaxy Clusters}

\author[a,b,1]{Nihar Dalal\note{Corresponding author.}}
\author[a,b,c,d]{Chun-Hao To}
\author[a,b,c]{Chris Hirata}
\author[e]{Tae Hyeon-Shin}
\author[f,g]{Matt Hilton}
\author[h]{Shivam Pandey}
\author[i]{J. Richard Bond}

\affiliation[a]{ Center for Cosmology and AstroParticle Physics (CCAPP), The Ohio State University, Columbus, OH 43210, USA}
\affiliation[b]{  Department of Physics, The Ohio State University, Columbus, OH 43210, USA}
\affiliation[c]{  Department of Astronomy, The Ohio State University, Columbus, OH 43210, USA}
\affiliation[d]{Department of Astronomy and Astrophysics, University of Chicago, Chicago, IL 60637, USA}
\affiliation[e]{ Department of Physics and Astronomy, Carnegie Mellon University, Pittsburgh, PA 15213, USA}
\affiliation[f]{Wits Centre for Astrophysics, School of Physics, University of the Witwatersrand, Private Bag 3, 2050, Johannesburg, South Africa}
\affiliation[g]{Astrophysics Research Centre, School of Mathematics, Statistics, and Computer Science, University of KwaZulu-Natal, Westville Campus, Durban 4041, South Africa}
\affiliation[h]{William H. Miller III Department of Physics \& Astronomy, Johns Hopkins University, Baltimore, MD 21218, USA }
\affiliation[i]{Canadian Institute for Theoretical Astrophysics, University of Toronto, ON M5S 3H8, Canada}

\emailAdd{dalal.64@osu.edu}

\abstract{The next generation of cosmology surveys will probe the matter distribution of the universe to unparalleled precision. To match this level of precision in cosmological parameter estimation, we need to use information at small scales of $\sim$ 1 Mpc, which requires an accurate model of baryonic feedback.  In this paper, we employ the Dark Matter + Baryon (DMB) model, a flexible halo model that is well-fit to various hydrodynamical simulations, to extract information on baryonic feedback from galaxy cluster observables. Using a sample of thermal Sunyaev-Zeldovich (tSZ) selected galaxy clusters from the Atacama Cosmology Telescope (ACT) - with masses calibrated via weak lensing from the Dark Energy Survey (DES) - we develop a robust end-to-end pipeline that directly models the calibrated observables. Our analysis demonstrates that the tSZ Y-M relation can constrain several DMB model parameters, providing key insights into baryonic feedback effects on cosmic shear at the several percent level. We find a preference for intermediate to strong levels of feedback, which is both consistent with several hydrodynamic simulations and competitive with similar analyses performed on complementary probes. Finally, we discuss the implications of our results in the context of current and upcoming cosmic shear surveys.}

\begin{document}
\maketitle
\flushbottom

\section{Introduction}
\label{sec:intro}
Weak gravitational lensing probes the total matter distribution in our universe, providing insight into the formation of structure in the universe, independent of bias models that relate the dark matter to directly observable tracers \cite{Mandelbaum:2017jpr}. While gravity drives the growth of structure, smaller-scale astrophysical processes such as feedback from Active Galactic Nuclei (AGN) alter the distribution of baryonic matter. Galaxy clusters, as the most massive collapsed structures, are the hotbeds of this energetic physics, making them critical laboratories for studying both the growth of structure and baryonic feedback.

In the era of precision cosmology, baryonic feedback has gained newfound importance as it causes a noticeable suppression in the measured matter power spectrum at the level of up to 30\% \cite{Chisari2019} at scales of $k>1~h\rm{Mpc}^{-1}$, which are well probed by current and future weak lensing surveys. Unbiased inference of cosmological parameters from weak lensing requires an accurate model of this suppression. One set of approaches has been to calibrate the suppression with hydrodynamic simulations \cite[see][for a recent example]{FlamingoSupp}. A drawback of these approaches is that simulations are computationally expensive, and the feedback is sensitive to prescriptions of subgrid physics \cite{2020NatRP...2...42V}.  
A more conservative approach adopted by cosmology analyses from surveys including the Dark Energy Survey (DES) \cite{Y3_Secco, Y3_Amon, DESY3, DESY1}, Hyper Suprime Cam (HSC) \cite{HSC}, and Kilo Degree Survey (KiDS) \cite{KIDScombine}, is to remove the signal from scales where modeling uncertainties dominate by enforcing scale cuts. While making the analysis more robust, this approach will forgo lots of signal-to-noise measured by future instruments \cite{Eifler:2024jai}. 

Semi-empirical models (see \cite{Schneider19} and \cite{Arico:2020lhq}) present a promising middle ground, parameterizing baryon-induced modifications to dark matter halos in a flexible and computationally efficient manner. However, constraining these models requires observational anchors. Several studies have approached these constraints by jointly modeling several probes. For example, \cite{ACTxDEStsZ1, ACTxDEStsZ2} looked at cross-correlations between ACT tSZ maps and DES weak lensing (WL) shear at the 2pt-statistic level to constrain pressure profiles in halos. Studies such as \cite{Grandis2023, Giri21} have looked at jointly modeling WL and X-ray data, whereas \cite{DES:2024iny} considered using WL alongside the kSZ signal with velocity reconstructions from the BOSS CMASS galaxy sample. Other studies \cite{Arico2023, 2023MNRAS.518.5340C, 2024JCAP...08..024G} have also placed robust constraints on the suppression of the matter power spectrum from weak lensing data alone. 

Among different probes, galaxy clusters, with their dual role as significant contributors to the lensing power spectrum at feedback-sensitive scales \cite{Paper1} and as tracers of baryonic physics, present unique strengths. Their hot intracluster gas, for instance, imprints a thermal Sunyaev-Zeldovich (tSZ) signal on the CMB, providing a direct probe of gas thermodynamics that complements weak lensing. Furthermore, clusters are bright across several different wavelengths, which provide distinct information at different mass scales \cite{Weinberg2013}. Last but not least, the selection function of clusters is well understood, and therefore easier to model than luminous galaxies.

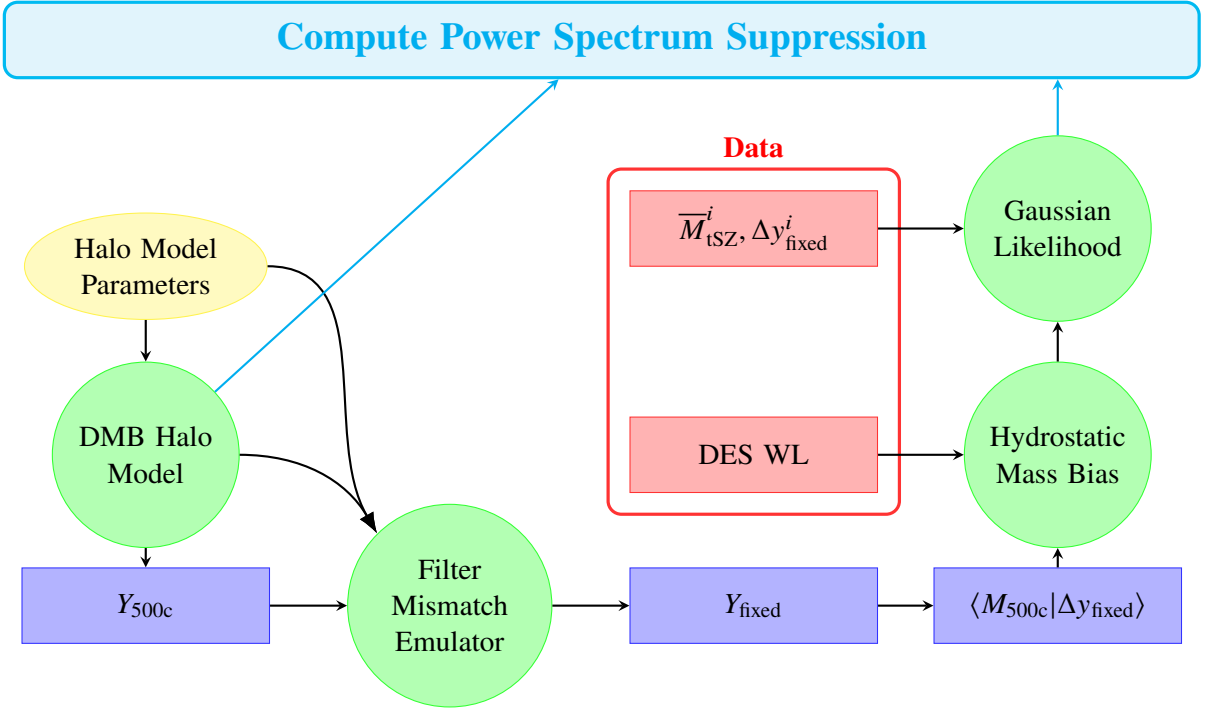
\begin{figure}
    \centering
    \begin{tikzpicture}[node distance=2cm]
    \usetikzlibrary{shapes, arrows.meta, fit} 
    
    \node (n1) [input] {Halo Model Parameters};
    \node (dmb) [model, below of=n1, yshift=-0.5cm] {DMB Halo Model};

    \node (constrain) [rectangle, 
    draw = cyan!80, 
    fill = cyan!10, 
    very thick,
    right of=n1, 
    xshift = 4cm,
    yshift = 3cm,
    minimum width =1cm,
    minimum height = 1cm, 
    text width =  \linewidth,
    text = cyan,
    rounded corners = 5pt,
    font=\Large\bfseries, 
    align=center
    ] {Compute Power Spectrum Suppression};
    
    \node (n2) [process, below of=dmb] {$Y_{\rm{500c}}$};
    \node (emu) [model, right of=n2, xshift=2cm] {Filter Mismatch Emulator};
    \node (n3) [process, right of=emu, xshift=2cm] {$Y_{\rm{fixed}}$};
    \node (n4) [process, right of=n3, xshift=2cm] {$\expval{M_{\rm{500c}}| \Delta y_{\rm{fixed}}}$};
    \node(likelihood)[model, above of=n4, yshift = 1cm]{Gaussian Likelihood};
    \node (n5) [rectangle, minimum width=1cm, minimum height=1cm, text centered, 
        text width=3cm, draw = red!80, fill = red!30, xshift = -2cm, left of=likelihood] {$\overline{M}_{\rm{tSZ}}^i, \Delta y_{\rm{fixed}}^i$};
    \node(WL)[rectangle,minimum width=1cm, minimum height=1cm, text centered, above of =n5, text width=3cm, draw = red!80, fill = red!30] {DES WL};

     \node[draw=red!80, very thick, rounded corners=5pt,
          fit=(WL)(n5), inner sep=8pt,
          label={[text=red, font=\bfseries]above:Data}] (databox) {};

    \draw[arrow] (n1) -- (dmb);
    \draw[arrow] (dmb) -- (n2);
    \draw[arrow] (n2) -- (emu);
    \draw[arrow] (emu) -- (n3);
    \draw[arrow] (n3) -- (n4);
    \draw[arrow] (n5) -- (likelihood);
    \draw[arrow] (n4) -- (likelihood);
    \draw[arrow, cyan] (likelihood.north) -- ([xshift=6cm]constrain.south);

    \draw[arrow, -{Latex[scale=1.2]}] 
        (dmb.east) to [out=0, in=120] (emu.north west);
    \draw[arrow, -{Latex[scale=1.2]}] 
        (n1.east) to [out=0, in=120] (emu.north west);
    \draw[arrow, cyan] (dmb)--(constrain);
\end{tikzpicture}
    \caption{A schematic for the modeling approach that we take in this paper. Input parameters are given by the yellow ellipse and the data are denoted by red rectangles, whereas computed quantities are in blue rectangles. The green circles detail the various modeling components that are used to describe the data vector, and our main result is the computed suppression in the power spectrum. }
    \label{fig:process}
\end{figure}

In this paper, we carry out the approach outlined in \cite{Paper1}, which offers a pathway to simultaneously constrain feedback models and self-calibrate small-scale systematics for next-generation cosmic shear surveys.  We employ the clusters detected in the Atacama Cosmology Telescope (ACT) DR5 \cite{Hilton_2021}, whose mass is calibrated with the weak lensing data measured in the Dark Energy Survey (DES) Year 3 \cite{DESY3Shape}. We provide a rigorous characterization of the instrument and its cluster detection pipeline. Specifically, noise-dominated tSZ clusters have traditionally been detected with a series of matched filters, and the detection process is highly sensitive to the true signal and shape of the cluster. We therefore develop a robust emulator to characterize the filter mismatch function as it varies with model parameters, which allows us to fully forward model the mass-observable relationship from ACT data. With this well-calibrated observable, we are able to provide observationally motivated constraints on the suppression of the matter power spectrum. A schematic of our modeling approach is presented in Figure \ref{fig:process}, where we attempt to characterize the various modeling components that enter our analysis. 

While there are many important details in our analysis, the underlying principle is straightforward. Stronger baryonic feedback leads to a more diffuse baryon distribution in and around massive halos, thereby reducing the tSZ signal at a fixed halo mass. At fixed tSZ signal, conversely, the measured halo mass is larger if feedback is stronger. From weak-lensing calibrated measurements of the mean halo mass as a function of the tSZ signal, we can reasonably well constrain our parametrized feedback model to determine the impact of feedback on the matter power spectrum.  

The paper is organized as follows. In section~\ref{sec:theory}, we outline the Dark Matter + Baryon (DMB) halo model and describe the relevant parameters. We also detail the modeling of the tSZ signal and the matter power spectrum. In section~\ref{sec:data}, we describe the cluster catalog and mass calibration that is used in our study. In section~\ref{sec:method}, we outline the modeling process, including our filter mismatch emulator. In section~\ref{sec:results}, we present our constraints on DMB parameters and discuss the implications for baryonic suppression.  We provide a conclusion in section~\ref{sec:conc}.

\section{Theory Framework}
\label{sec:theory}
\subsection{A DMB halo model}
\label{sec:1h}
This section delineates our implementation of the DMB halo model, which was originally presented by \cite{Schneider15}, and subsequently modified by \cite{Schneider19, methodpaper, Paper1, 2024OJAp....7E.108A}. A complementary approach to model both the matter distribution and the tSZ signal has been developed upon the Bacco model by \cite{2024A&A...690A.188A}.  
We model the matter distribution $\rho(r,M_{200c},z, c_{200c})$ around an isolated halo at a given mass ($M_{200c}$), concentration ($c_{200c}$) and redshift ($z$)  with three components, namely
\begin{eqnarray}
    \rho(r,M_{200c},z, c_{200c}) = \rho_{\rm{gas}} (r,M_{200c},z)+\rho_{\rm{cga}} (r,M_{200c})+\rho_{\rm{clm}}(r,M_{200c},z, c_{200c}), 
\end{eqnarray}
where  $\rho_{\rm{gas}} (r,M_{200c},z)$ describes the intracluster medium, $\rho_{\rm{cga}} (r,M_{200c})$ describes stars in the central galaxy of the halo, $\rho_{\rm{clm}} (r,M_{200c},z, c_{200c})$ consists of stars in satellite galaxies and dark matter, and $r$ is the comoving distance to the halo center. The radius $r_{200c}$ is defined as the radius of the sphere containing matter that has average density $200 \rho_{c}$, where $\rho_c$ is the critical density of the universe. The mass $M_{200c}$ and the concentration $c_{200c}$ are defined with respect to this radius.  

These components depend on the total stellar fraction, $f_{\rm{star}}$, and the stellar fraction in the central galaxy, $f_{\rm{cga}}$, whose dependence on halo mass is modeled following \cite{2013MNRAS.428.3121M}: 
\begin{eqnarray}
\label{eq:star}
f_{\rm{star}} (M_{200c}) = 0.055 \left(\frac{2.5\times10^{11}~ h^{-1}M_\odot}{M_{200c}}\right)^{\eta_*}, \\
\label{eq:cga}
f_{ \rm{cga}} (M_{200c}) = 0.055 \left(\frac{2.5\times10^{11}~ h^{-1}M_\odot}{M_{200c}}\right)^{\eta_{ \rm{cga}}},
\end{eqnarray}
where $\eta_{\rm{cga}} = \eta_*+\delta\eta$.

The gas component $\rho_{\rm{gas}} (r,M_{200c},z)$ is parametrized by a cored double power law \cite{Schneider19, Giri21}
\begin{equation}
\label{eq:rhogas}
    \rho_{\rm{gas}} (r,M_{200c},z) \propto \frac{\Omega_{\rm{b}}/\Omega_{\rm{m}}-f_{\rm{star}}(M_{200c})}{\left(1+10 r/R_{200c}\right)^{\beta(M_{200c},z)} \left(1+ r/\left(R_{200c}\theta_{\rm{ej}}\right)\right)^{\left(\delta-\beta\left(M_{200c},z\right)\right)/\gamma}},
\end{equation}
where $\theta_{\rm{ej}}, \delta$, and $\gamma$ are free parameters describing the gas ejection radius and slope, and $\beta (M_{200c},z)$ is modeled as 
\begin{equation}
\label{eq:betam}
    \beta (M_{200c},z) = \frac{3}{1+(M_{c0}(1+z)^{\nu_{\rm z}}/M_{200c})^{\mu_{\beta}}}, 
\end{equation}
 where $M_{c0}$, $\nu_{\rm z}$, and ${\mu_{\beta}}$ are free parameters.

The central galaxy component described a truncated Gaussian profile \cite{Schneider19, Giri21}:
\begin{equation}
\label{eq:cg}
    \rho_{\rm{cga}} (r,M_{200c})=\frac{f_{\rm{cga}}(M_{200c})}{0.06\pi^{3/2}R_{200c}r^2} \exp \left(-\left(\frac{r}{0.03 R_{200c}}\right)^2\right).
\end{equation} 

In a gravity only simulation, collisionless matter should follow a truncated NFW profile \cite{nfw, 2011MNRAS.414.1851O}, albeit modified by the gravitational potential of the gas and central galaxy components. The timescale of the change in potential is assumed to be significantly longer than the dark matter's orbital period, so we can model the evolution as an adiabatic process. Hence, $\rho_{\rm{clm}}$ is given by \cite{Schneider19}
\begin{eqnarray}
\label{eq:clm}
    \rho_{\rm{clm}}(r,M_{200c},z, c_{200c}) &=& \left(1-\Omega_{\rm{b}}/\Omega_{\rm{m}}+f_{\rm{star}}(M_{200c})-f_{\rm{cg}}(M_{200c})\right) \rho_{\rm{nfw}}\left(r/\zeta(r,M_{200c},z, c_{200c}), c_{200c}\right)\nonumber\\
    \rho_{\rm{nfw}}(r, c_{200c})&\propto& 
    \frac{1}{\left(\frac{rc_{200c}}{R_{200c}}\right)\left(1+\frac{rc_{200c}}{R_{200c}}\right)^2} \frac{1}{\left(1+\left(\frac{r}{4R_{200c}}\right)^2\right)^2},
\end{eqnarray}
where $R_{200c}$ is the halo radius. We can hence solve for $\zeta (r,M_{200c},z, c_{200c})$ following \cite{2010MNRAS.407..435A, 2011MNRAS.414..195T} by assuming the initial radius $r_i$ and final radius $r_f$ of a shell are related by
\begin{equation}
\begin{aligned}
    \zeta -1 &= \frac{r_f}{r_i} - 1 = 0.3 \left[ \left(\frac{M_i}{M_f}\right)^2 -1 \right]\\&=0.3\left(\left(\frac{\int_0^r\rho_{\rm{nfw}}(t,c_{200c}) t^2 dt}{\int_0^r \rho_{\rm{clm}}(t, M_{200c}, z, c_{200c}) t^2 dt + \int_0^{\zeta r} \left(\rho_{\rm{gas}}(t,M_{200c},z)+\rho_{\rm{cg}}(t,M_{200c}) \right)t^2 dt}\right)^2-1\right).\\
\end{aligned}
\end{equation}
Finally, $\rho_{\rm{clm}}(r,M_{200c},z, c_{200c})$ and $\rho_{\rm{gas}}(r,M_{200c},z)$ are normalized by the same factor such that 
\begin{eqnarray}
    \int_0^\infty \left(\rho_{\rm{gas}}(r)+\rho_{\rm{cga}}(r)+\rho_{\rm{clm}}(r)\right) 4\pi r^2 dr = \int_0^\infty \rho_{\rm{nfw}}(r) 4\pi r^2 dr.
\end{eqnarray}
\begin{figure}[htbp]
\centering
\includegraphics[width=\textwidth]{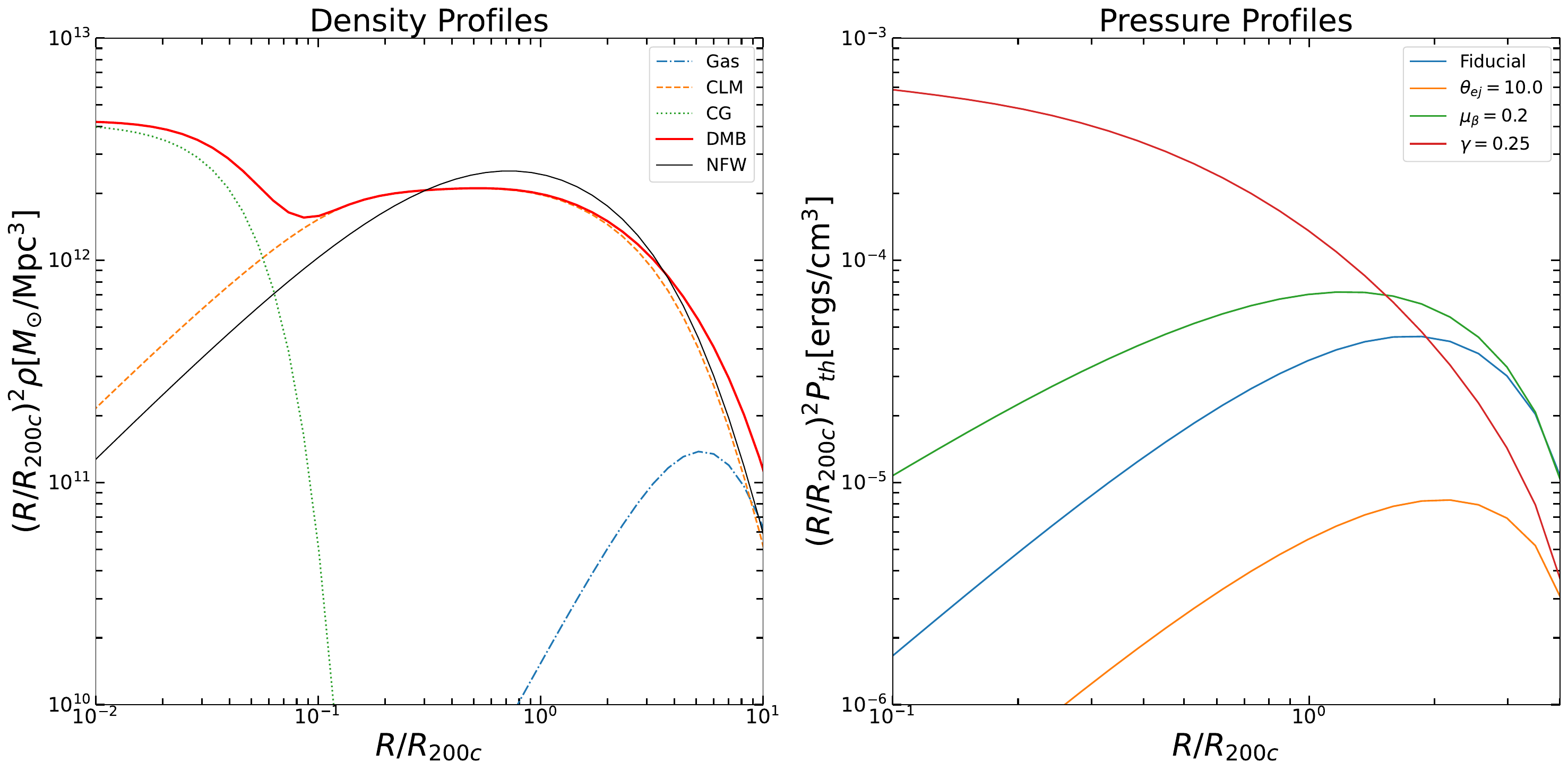}
\caption{Left: A plot of the different components of the DMB halo model (red) as compared to the a standard NFW profile (black). The blue dash-dotted line is the gas component of the profile described by Equation \ref{eq:rhogas}, the orange dashed line is the collisionless matter component described by Equation \ref{eq:clm}, and the green dotted line is the central galaxy component given by Equation \ref{eq:cg}. Right: The thermal pressure profile as it varies for different choices of DMB model parameters. In blue, we have the fiducial values, which are given by the mean of the prior, as shown in Table \ref{tab:Halomodel}. In orange, we vary the extent of the gas profile with $\theta_{\rm{ej}}$; in green and red we vary the gas slope by changing $\mu_\beta$ and $\gamma$ respectively.  \label{fig:i}}
\end{figure}

\begin{table*}
\begin{tabular}{l l l l}
\hline
\hline
Parameter  & Prior  & Description & Relevant equations \\
\hline
\multicolumn{3}{c}{\textbf{Gas component}} \\ 

$\log_{10} M_{c0}$ & flat(13, 15) & \multicolumn{1}{l}{\multirow{2}{9cm}{The pivot mass scales when the slope of gas profile becomes shallower than $3$.}} & Equation \ref{eq:betam}\\\\

$\nu_{\rm z}$ & flat(-1,1) & Redshift dependence of the pivot mass scale. & Equation \ref{eq:betam}\\

${\mu_{\beta}}$ & flat(0,2) & Mass dependence of profile slope.& Equation \ref{eq:betam}\\

$\theta_{\rm{ej}}$ & flat(2,8) & Maximum radius of gas ejection relative to $R_{200c}$. & Equation \ref{eq:rhogas}\\

$\gamma$ & flat(1,4) & Slope of gas profile. & Equation \ref{eq:rhogas}\\

$\delta$ & flat(3,11) & Slope of gas profile. & Equation \ref{eq:rhogas}\\

$\alpha_{\rm{nt}}$ & flat(0.01, 0.4) & Fraction of gas pressure caused by non-thermal processes. & Equation  \ref{eq:thermal}\\ 
$n_{\rm{nt}}$ & flat(0.6,1.0) & Redshift evolution of non-thermal gas pressure fraction. & Equation \ref{eq:thermal}\\
\hline
\multicolumn{3}{c}{\textbf{Stellar component}} \\
$\eta_*$ & flat(0.15,0.3) & Mass dependence of stellar fraction. & Equation  \ref{eq:star}\\ 

$\delta \eta$ & flat(0.05,0.4) & {Difference in mass dependence of stellar fraction} & Equation  \ref{eq:cga}\\
\hline
\multicolumn{3}{c}{\textbf{Y-M relation}} \\
$\sigma_{\ln y}$ & flat(0.1, 0.3) & Scatter of the Y-M relation & Equation \ref{eq:lognormal}
\\
\hline 
\hline 
\end{tabular}
\centering
\caption{\label{tab:Halomodel} Parameters and priors of our halo model described in section \ref{sec:1h}.} 
\end{table*}

\subsection{Computing the tSZ Signal}
The modeling of the tSZ signal follows from \cite{methodpaper, Paper1}. Assuming hydrostatic equilibrium, we find the total pressure profile $P_{\rm{tot}}(r)$ from the gas density profile:
\begin{eqnarray}
    P_{\rm{tot}}(r) = \int_r^\infty \frac{GM_{\rm{tot}} (<t)}{t^2}\rho_{\rm{gas}}(t, M_{200c},z) dt, \\
    M_{\rm{tot}}(<r) = 4\pi \int_0^r t^2\rho(t, M_{200c},z) dt.
\end{eqnarray}
The total pressure consists of a thermal component, $P_{\rm{thermal}}$ and contributions from non-thermal processes that do not impact the tSZ signal. Following \cite{2014ApJ...792...25N}, we use a simple fitting formula to determine the fraction of thermal pressure with another set of free parameters $\alpha_{\rm{nt}}$ and $n_{\rm{nt}}$:
\begin{eqnarray}
\label{eq:thermal}
    \frac{P_{\rm{thermal}}}{P_{\rm{tot}}}&=& {\rm{ max}} [0, 1-\alpha_{\rm{nt}} f(z) (r/R_{500c})^{0.8}]\nonumber, \\
    f(z) &=& {\rm{min}} [(1+z)^{0.5}, (4^{-n_{\rm{nt}}}/\alpha_{\rm{nt}}-1){\rm{tanh}}(0.5z)+1]
\end{eqnarray}
Under the final key assumptions that gas is fully ionized and electrons and protons are in thermal equilibrium within galaxy clusters, we can model the electron pressure $P_{e}$ as
\begin{equation}
    P_e = \frac{4-2Y}{8-5Y}P_{\rm{thermal}} 
\end{equation} 
where $Y = 0.24$ is the primordial helium mass fraction. The equilibrium assumption is valid at the centers of galaxy clusters, but it may break down at the outskirts \cite{2009ApJ...701L..16R, 2015ApJ...808..176A}. All quantities that we compute and use in the analysis are taken with respect to $R_{\rm{500c}}$, which is well within the region where our equilibrium assumption is valid. 

The integrated tSZ signal $y_{\rm{500c}}$ for a cluster within radius $R_{\rm{500c}}$ can therefore be computed as 
\begin{equation}
\label{eq:integratedy}
    y_{\rm{500c}} = \frac{\sigma_T}{m_ec^2} \int_0^{R_{500c}} P_e(r) 4\pi r^2 \, dr,
\end{equation}
where $\sigma_{\rm{T}}$ is the Thomson cross section, $m_{\rm{e}}$ is the electron mass, and $c$ is the speed of light. The mean mass of clusters given $y_{\rm{500c}}$ can then be computed assuming a log-normal $y_{\rm{500c}}-M$ relation, with mean given by \ref{eq:integratedy}, and scatter $\sigma_{\ln y}$ as a free parameter. 

The 10 free parameters of this model and the associated priors assumed in analyses presented in this paper are summarized in Table \ref{tab:Halomodel}, along with the priors we use for our analysis. Priors on the stellar parameters, namely $\eta_*$ and $\delta \eta$, are obtained from previous studies including \cite{Giri21} which place constraints on parameters from observed gas and stellar fractions from both X-ray and optical observations of galaxy clusters. Given the complexity of the model and the multitude of parameters, we refer readers to Appendix \ref{sec:ParameterDependence} for a more intuitive understanding  of how certain parameters affect both the observable $y$-$M$ relation and the suppression in the matter power spectrum.

\section{Data}
\label{sec:data}

\subsection{Atacama Cosmology Telescope}
The Atacama Cosmology Telescope (ACT) \cite{ACTInstrument} saw first light in 2007 on Cerro Toco in the Atacama Desert in northern Chile. The 6-meter, millimeter wavebandm ground based CMB experiment ran until 2022, undergoing two major receiver upgrades: ACTPol (2013 - 2016) \cite{ACTPolInstrument}, and Advanced ACTPol (2017-2022) \cite{advactpol1, advactpol2}. Precise CMB maps made from ACT data have led to numerous science outcomes including several tSZ cluster catalogs. 

We use the ACT DR5 cluster catalog \cite{Hilton_2021}, which contains 4195 optically confirmed tSZ clusters detected with signal to noise $> 4$ spanning over 13,000 $\rm{deg}^2$ of the sky. We restrict the cluster catalog to the
redshift range where the mass bias relation (see \ref{subsec:bias}) is valid, i.e. with clusters such that $0.15 \leq  z \leq 0.7$.
ACT tSZ clusters are detected using a multi-frequency matched filter applied to the 90 GHz and 150 GHz maps. The ACT collaboration models cluster signals using the Universal Pressure Profile (UPP) \cite{Arnaud_2010} and convolving with the appropriate ACT beam for a given frequency to form a signal template. A set of 16 matched filters is formed using this method \cite{Hilton_2021}. 

In addition to identifying and characterizing the cluster signal with the best matched filter, the catalog also characterizes the tSZ signal $y_{\rm{fixed}}$ and survey completeness with a fixed reference filter scale of $\theta_{500c} = 2.4'$, corresponding to a UPP-model cluster at $z = 0.4$ with mass $M_{500c} = 2 \times 10^{14} M_{\odot}$ at the fiducial cosmology. 

The advantage of using a fixed filter as opposed to a varying filter scale with every cluster is the removal of inter-filter noise bias \cite{Hasselfield2013}. In particular, the preferred filter scale for a given cluster is equally affected by both the true signal from the cluster and the amplitude of local noise. When characterizing the signal with a single filter scale, the primary CMB signal acts as a source of Gaussian noise for the desired cluster signal. In the context of cosmological results from the ACT cluster catalogs, the choice of using a fixed filter scale simplifies the selection function, which is crucial for determining and characterizing mass observable relationships that can be used for cosmological inference.

Given that ACT maps over 18,000 square degrees, we also expect the noise level to vary considerably as a function of position on the sky. Following the method outlined in \cite{Naess:2020wgi,Hilton_2021}, the maps are split into 280 equal-sized, overlapping tiles of size $5^\circ \times 10^\circ$ with reasonably constant levels of white noise. Given this variance in noise, a different matched filter is constructed for each individual tile. We attempt to average over this noise variance over tiles, as we detail further in Section \ref{sec:method}. 

\subsection{Mass calibration bias}
\label{subsec:bias}

The $M_{\rm{500c}}$ masses reported in the ACT catalog are derived from the Universal Pressure Profile scaling relation \cite{Arnaud_2010}, which assumes a hydrostatic equilibrium in computing the relation. True masses of these objects with calibration from WL data are often a factor of about $\sim 0.71$ lower, such as the richness-based cluster WL calibration from DES data, which is also reported in the ACT catalog \cite{Hilton_2021}. The true masses, and therefore the masses that should be used in our model are better estimated by weak lensing. Further, we expect this mass calibration bias to not be constant across cosmic time and different mass scales. \footnote{It might seem unnecessary to model $M_{\rm tSZ}$ and $b$ simultaneously instead of modeling the weak lensing profile. We note that this modeling approach is an appealing middle ground. While the halo model adopted and developed in this paper captures the necessary modeling ingredients to describe the impact of baryonic feedback on the tSZ signal, it fails to incorporate several key ingredients to model weak lensing profiles, such as mis-centering, photometric redshift uncertainties, and intrinsic alignments. While these effects might be correlated with the baryonic feedback signal, the shape noise in weak lensing profile is large enough that this correlation can be negligible \citep{SPT2}. On the other hand, the relationship between $M_{\rm tSZ}$ and the weak lensing profile is modeled in detail in Shin et al. (in preparation), where we can extract the $M_{\rm tSZ}$--$M_{\rm{500c}}$ relation after marginalized several aforementioned systematic effects. In the regime where those effects are uncorrelated with the baryonic feedback signal, our two-step approach will lead to an unbiased result.} 


In this study, we parametrize the mass calibration bias $b_{\rm{tSZ}}$ as a function of mass and redshift as, 
\begin{equation}
\label{eq:hydrobias}
    b_{\rm{tSZ}}(M_{\rm{500c}}, z) = \frac{M_{\rm{tSZ}}}{M_{\rm{500c}}} = \left[\frac{1}{A_{\rm{mass}}} \left(\frac{M_{\rm{500c}}}{3 \times 10^{14} M_{\odot}} \right)^{-\eta} \left(\frac{1+z}{1.45}\right)^{-\zeta}\right]^{\frac{1}{1+\eta}}, 
\end{equation}
where $\eta$, $\zeta$, and $A_{\rm{mass}}$ are free parameters. 

We constrain the mass bias model using the weak-lensing technique, with the data taken during the first three years of the Dark Energy Survey (DES-Y3) \cite{DESY3Shape}. While we refer the reader to Shin et al. (in prep.), we briefly describe our methodology here.

First, we measure the stacked tangential shear ($\gamma_{\rm t}$) around the clusters as a function of angular distance ($\theta$). We split our cluster sample into two redshift bins and two bins of $y_{\rm fixed}$, each of which has approximately the same number of objects. For the weak-lensing source galaxies, we remove the first two source redshift bins (out of four), since they overlap with or lie in front of our cluster sample in redshift space, and we use the remaining source galaxies to measure our tangential shear data vector ($\gamma(\theta)$). The boost factor\footnote{The contamination of the tangential shear signal due to the cluster member galaxies leaked into the source galaxy sample.} is measured using the $P(z)$ decomposition method \cite{Gruen2014,Varga2019}. 

Instead of correcting the tangential shear data vector with the measured boost factor, we instead jointly fit the measured tangential shear and the boost factor, accounting for the hydrostatic mass bias as modeled above. The boost factor is modeled following \cite{Tomclusterlensing}, while the stacked tangential shear is modeled as the sum of a projected NFW profile \cite[][dark matter contribution]{nfw} and a generalized NFW (gNFW) profile \cite[][baryonic contribution]{Zhao1996}. We refer the readers to \cite{Cromer2022} for details of this model. The final model for the stack tangential shear is derived by taking the average of the models of the individual clusters:
\begin{eqnarray}\label{eq:gt_mean_sys}
\hat{\boldsymbol{g}}_{t,j} = [(1+m_{\gamma,j}) / \mathcal{B}_j] \sum_i w_i \hat{\boldsymbol{g}}_{t,ij},
\end{eqnarray}
where $j$ is the index for the source redshift bin, $1+m_{\gamma,j}$ the shear calibration bias of the $j$-th source redshift bin \cite{MacCrann2022}, $\mathcal{B}_j$ the boost factor model for the $j$-th source redshift bin, $w_i$ the total weight of the source galaxies used for the $i$-th cluster to account for the different lensing weights of each cluster in the data vector, and finally $\hat{\boldsymbol{g}}_{t,ij}$ the tangential shear model for the $i$-th cluster using $j$-th source redshift bin. Note that the hydrostatic mass bias, $b(M_{\rm{500c}}, z)$, is applied to the individual cluster model, $\hat{\boldsymbol{g}}_{t,ij}$. Since the weak-lensing mass calibration is not a main focus of this study, we refer the readers to Shin et al. (in prep.) for details of this model. For future analyses, we hope to unify the model used in this work and the work used in Shin et al. (in prep.) for consistency. 

Given the model for the stacked tangential shear above, we perform an MCMC analysis to constrain the parameters in the hydrostatic mass bias ($A_{\rm mass}$, $\eta$, $\zeta$). We refer the readers to Shin et al. (in prep.) for details of likelihoods and running chains. We then propagate the posteriors of this analysis through our model as informative priors for the hydrostatic mass bias using normalizing flows \cite{normFlows}. While the bias parameters are allowed to vary within the prior ranges in our final analysis, we find the tSZ $y$-$M$ relation carries no new information about the hydrostatic bias parameters, and the posteriors for these parameters are purely prior dominated. The final posteriors are shown in the last three panels of \ref{fig:ii} in Appendix \ref{sec:MCMC}. 

One might also wonder about the effects of baryonic feedback on the WL mass profiles, and therefore the effect that mismodeling baryons may have on the mass calibration bias. We point to the results of \cite{Cromer2022}, where the authors find using a combination of NFW and gNFW profiles has at most a 1\% bias in mass estimations compared to the true masses measured in hydrodynamic simulations. Ideally, we would jointly model both the WL profile and the tSZ signal with our DMB halo model, but we leave this for a future study. We expect the difference between using our DMB model and the NFW + gNFW approach of Shin et al to therefore also be $\sim$ 1\%, given that the DMB halo model describes hydrodynamic simulations accurately. This level of bias is beyond the scope of precision of this work. Furthermore, we once again stress that our data contains no new information about the mass calibration bias, and therefore, we can use the results of Shin et al. as an external, unbiased prior.   
\subsection{Data Vector}
\begin{figure}
    \centering
    \includegraphics[width=\linewidth]{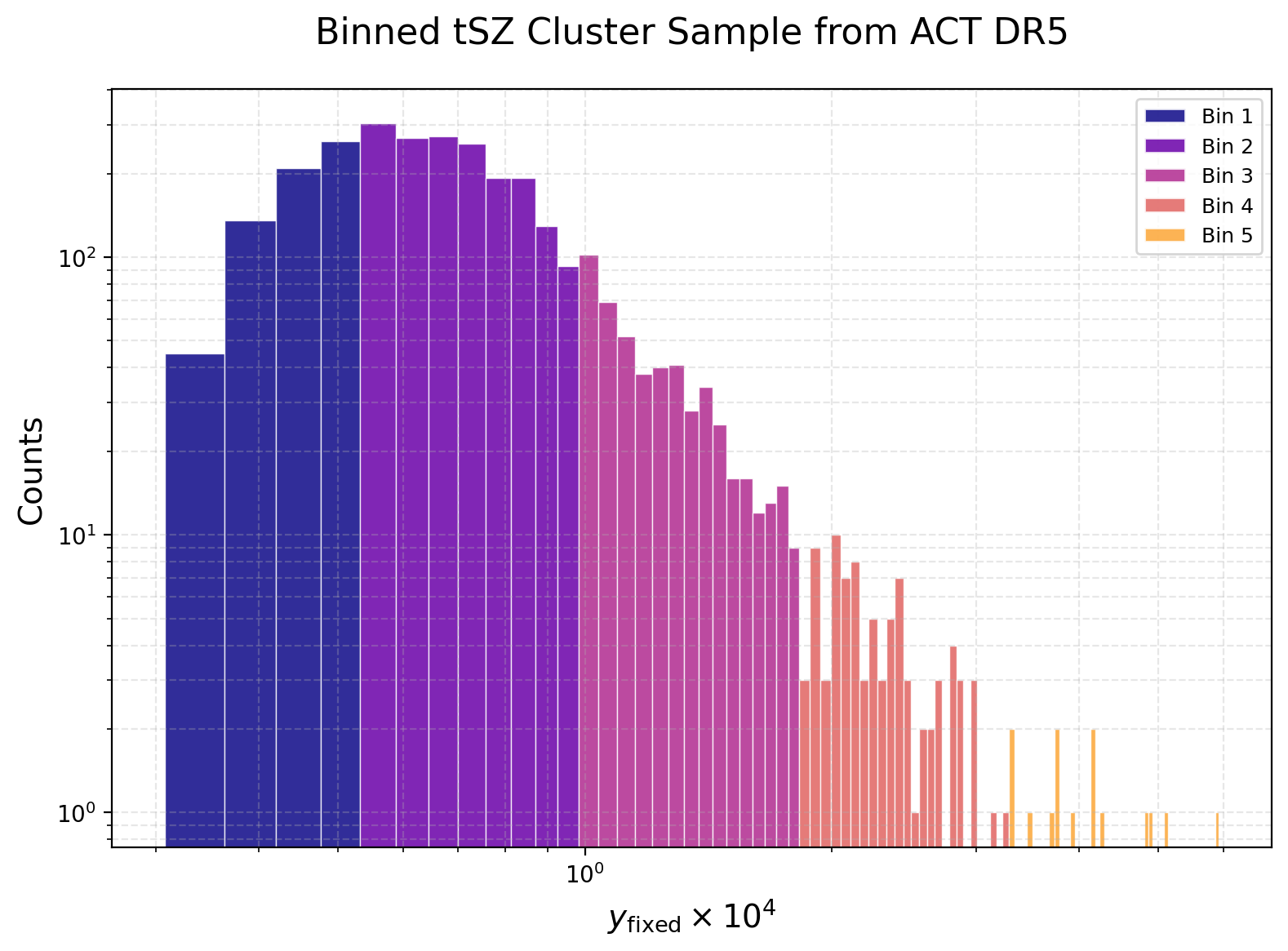}
    \caption{Histogram of the ACT DR5 cluster catalog in logarithmic bins of $y_{\rm{fixed}}$ (\texttt{fixed\_y\_c} in the catalog). Colors indicate the five larger bins that we use in our analysis.}
    \label{fig:bins}
\end{figure}

For the data vector, we choose to construct 5 logarithmic bins in $y_{\rm{fixed}}$, of widths $\Delta y_{\rm{fixed}}^i$ and we compute both a mean tSZ mass, $\overline{M}_{\rm{tSZ}}^i$, and an error on the mean, $\sigma_{\overline{M}_{\rm{tSZ}}}^i$, via bootstrapping. We choose logarithmic bins because of the long tail of the cluster catalog towards the high $y_{\rm{fixed}}$ end. Furthermore, we are relatively agnostic to the number of bins given that we integrate over $y_{\rm{fixed}}$ in modeling the data vector. Our bins are shown in Figure \ref{fig:bins}. 

\section{Method}
\label{sec:method}
\subsection{Filter Mismatch Function Emulator}
In this study, we vary the actual profile shapes of clusters using the DMB halo model. It is important to recognize that the computed $Y_{\rm{500c}}$ is not a direct observable. Furthermore, we expect the filter response to vary with the new shapes and profiles of the clusters. A crucial step in our method is therefore emulating the matched filter process itself as a function of DMB model parameters, using a filter mismatch function $Q$ described below. 

In previous studies \cite{Hilton_2021, Hasselfield2013}, the filter mismatch function has been the ratio of the fixed filter measurement (with $\theta_{\rm{500c}} = 2.4'$) of the central Compton Y parameter integrated along the line of sight, $y_{\rm{fixed}}$ (\verb|fixed_y_c| in the catalog), to the central Compton Y parameter of a cluster of potentially different angular size that was passed through the filter, $y_{\rm{true}}$ (\verb|y_c| in the catalog). We choose instead to recast the filter mismatch function as 
\begin{equation}
    Q(\theta_{\rm{500c}}, N_i) = \frac{y_{\rm{fixed}} (\theta_{\rm{500c}}, N_i)}{y_{\rm{500c}}(\theta_{\rm{500c}})}
\end{equation}
where $\theta_{\rm{500c}}$ refers to the angular size of the cluster, and $N_i$ refers to a noise tile. 
We constructed mock catalogs of galaxy clusters by drawing from the WebSky \cite{2020JCAP...10..012S} halo catalogs for a Latin Hypercube \cite{latinHyperCube} of DMB parameters in 10 dimensional space, and passed the catalogs through a modified version of the \textsc{Nemo} cluster detection package \cite{Hilton_2021}. We first painted the true signals of the clusters according to the DMB model in 2D, added the true CMB signal and inverse variance noise from \cite{Naess:2020wgi}, convolved with the ACT beam and fit template matched filters assuming the Universal Pressure Profile \cite{Arnaud_2010}. We also added the $1/f$ noise of the instrument to both the 90 $\rm{GHz}$ and 150 $\rm{GHz}$ maps as detailed in the appendix of \cite{2024ApJ...966..138M}. Finally, we characterized the filter mismatch by measuring a reported $Q(\theta, N_i)$, where $\theta$ refers to the angular size of the observed cluster, and $N_i$ refers to a noise tile. We then averaged $Q$ over area-weighted noise tiles,
\begin{equation}
\label{eq:filter}
    Q(\theta) = \frac{\sum_{N_i} Q(\theta, N_i) \cdot A (N_i)}{\sum A(N_i)},
\end{equation}
to retrieve a noise-weighted angular filter response function $Q(\theta)$ for a given set of DMB parameters. We then trained a chaos polynomial emulator to interpolate $Q(\theta)$ between points on the Latin Hypercube using the \textsc{Chaospy} \cite{chaospy} python package. 
\begin{figure}[htbp]
\centering
\includegraphics[width=\textwidth]{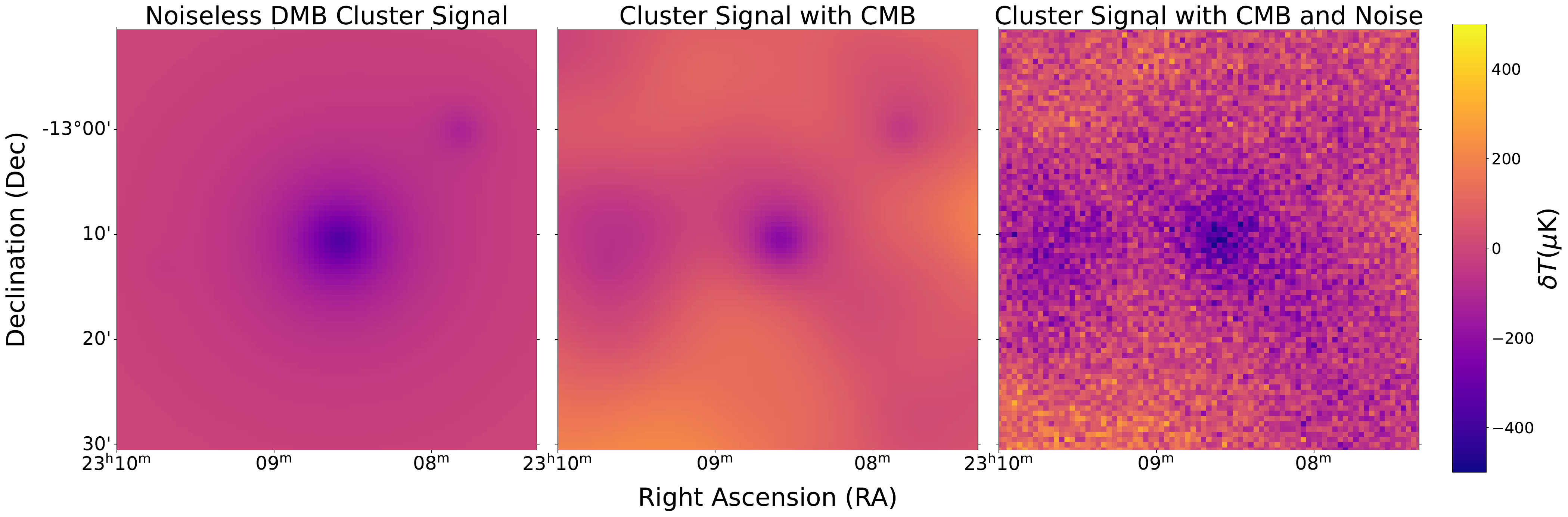}
\caption{Sample images taken from the Nemo cluster detection pipeline. The left panel shows the true DMB cluster signal as a change in temperature $\delta T$ in $\mu \rm{K}$ with a 90 GHz beam. The middle panel shows the cluster signal combined with a simulated primary CMB signal. The rightmost panel shows the impact of noise on the cluster signal, and the necessity for matched filters to identify the underlying cluster. \label{fig:iii}}
\end{figure}

Our choice to recast $Q$ as a ratio of the central Compton $Y$ parameter to the integrated Compton $Y$ parameter was intentional to make the model and pipeline more robust to issues of miscentering and optimization bias. In particular, drastically changing the profiles of clusters as normalized to their central $Y$ would significantly impact cluster detection, even if changes in DMB parameters were small. We therefore chose to instead normalize the profiles by $y_{\rm{500c}}$, which makes the cluster detection much more robust to changes in the profile, and more physically accurate. 
\subsection{Modeling}
 Our data vector is the mean tSZ mass $\overline{M}_{\rm{tSZ}} ^i$ computed in bins of widths $\Delta y_{\rm{fixed}}^i$, which we model as follows. We predict the expected ``tSZ'' or hydrostatic mass given a bin in $y_{\rm{fixed}}$ of some width $\Delta y_{\rm{fixed}}$ with
\begin{equation}
    \expval{M_{\rm{tSZ}}| \Delta \hat{y}_{\rm{fixed}}} =\frac{\int\int\int P(\hat{y}_{\rm{fixed}}|M_{\rm{500c}}, z) n(M_{\rm{500c}},z) M_{\rm{500c}} b_{\rm{tSZ}}(M_{\rm{500c}},z)\, \dd M_{\rm{500c}} \dd \hat{y}_{\rm{fixed}} \dd z}{\int\int\int P(\hat{y}_{\rm{fixed}}|M_{\rm{500c}}, z) n(M_{\rm{500c}},z) \, \dd M_{\rm{500c}} \dd \hat{y}_{\rm{fixed}} \dd z}
\end{equation}
where $b_{\rm{Hydro}}(M_{\rm{500c}},z)$ is given by \ref{eq:hydrobias}, $n(M_{\rm{500c}},z)$ is the Tinker halo mass function \cite{Tinker2010}, and $P(y_{\rm{fixed}}|M_{\rm{500c}})$ is given by a log-normal distribution, specifically:
\begin{equation}
\label{eq:lognormal}
    P(y_{\rm{fixed}}|M_{\rm{500c}}) = \frac{1}{\sqrt{2\pi} y_{\rm{fixed}} \sigma_{\ln y}} \exp \left(- \frac{(\ln y_{\rm{fixed}} - \ln y_{\rm{500c}}Q(\theta_{\rm{500c}}))^2}{2\sigma_{\ln y}^2} \right)
\end{equation}
with $Q$ as the noise-averaged filter mismatch function in Equation \ref{eq:filter}, $\sigma_{\ln y}$ is the scatter of the mass--observable relation, and $y_{\rm{500c}}$ is computed as a function of $M_{\rm{500c}}$ from the DMB halo model (see Equation \ref{eq:integratedy}).
We model the observed $\hat{y}_{\rm{fixed}}$ as drawn from a Gaussian distribution from the true $y_{\rm{fixed}}$ i.e.
\begin{equation}
    P(\hat{y}_{\rm{fixed}}|y_{\rm{fixed}}) = \mathcal{N}(y_{\rm{fixed}}, \sigma_{\rm{obs}}).
\end{equation}
and compute
\begin{equation}
    P(\hat{y}_{\rm{fixed}}|M_{\rm{500c}}) = \int P(\hat{y}_{\rm{fixed}}|y_{\rm{fixed}}) P(y_{\rm{fixed}}|M_{\rm{500c}}) \, \dd y_{\rm{fixed}}
\end{equation}
We finally fit the data as a function of DMB parameters $\theta_{\rm{DMB}}$ using a Gaussian likelihood, namely
\begin{equation}
    \log \mathcal{L}(\theta_{\rm{DMB}}) = -\frac{1}{2} \sum_{i=1}^5 \left(\frac{\expval{M_{\rm{tSZ}}| \Delta \hat{y}_{\rm{fixed}}^i} - \overline{M}^i_{\rm{tSZ}}}{\sigma^i_{\overline{M}_{\rm{tSZ}}}}\right)^2
\end{equation}
where $\overline{M}^i_{\rm{tSZ}}$ are taken from the ACT catalog, which uses the Universal Pressure Profile scaling relation \cite{Arnaud_2010} to compute the masses. 

\section{Results}
\label{sec:results}
\subsection{Computing the Matter Power Spectrum Suppression}
We can compute the matter power spectrum $P_{\rm{mm}}(k,z)$ analytically as the sum of the one-halo and two-halo terms. The one-halo term $P_{\rm{1H}}(k,z)$ is given by
\begin{equation}
    P_{\rm{1H}} (k,z) = \int_0^\infty \dd c_{\rm{200c}} \int_{M_1}^{M_2} W(M_{\rm{200c}}, k, z, c_{\rm{200c}})^2 n(M_{\rm{200c}}, z) p(c_{\rm{200c}}|M_{\rm{200c}}) \dd M_{\rm{200c}},
\end{equation}
where the halo concentration-mass relation $p(c_{\rm{200c}}|M_{\rm{200c}})$ is modeled as a log-normal distribution with scatter 0.11 dex\cite{2008MNRAS.390L..64D}, and a mean following \cite{2012MNRAS.423.3018P}, and lower mass $M_1$ and upper mass limits $M_2$ set to $10^{12}~h^{-1} M_\odot$ and $10^{16}~h^{-1} M_\odot$ respectively. The halo mass function $n(M_{\rm{200c}}, z)$ is modeled using the fitting function \cite{Tinker2010}, and $W(M_{\rm{200c}}, k, z, c_{\rm{200c}})$ is the fourier transform pair of $\rho(M_{\rm{200c}},r,z, c_{\rm{200c}})$ given by
\begin{eqnarray}
\label{eq:window}
    W(M_{200c},k,z,c) = \int_0^\infty 4\pi r^2 \frac{\sin(k r)}{kr}\frac{\rho(r, M_{200c}, z, c_{200c})}{\rho_{\rm m}(z)} dr,
\end{eqnarray}
where $\rho_m(z)$ is the mean matter density of the universe. 

The two halo term $P_{\rm{2H}}(k,z)$ is given by the product of the linear power spectrum and the square of the mean biases, namely
\begin{eqnarray}
\label{eq:salmon}
    P_{\rm{2H}}(k,z) &=& P_{\rm{lin}}(k,z) \left(I(M_1,k,z)+ A(M_1,z) \right)^2 \\
    I(M_1,k,z) &=& \int_0^\infty dc_{200c}\int_{M_1}^{\infty} W(M_{200c},k,z, c_{200c})n(M_{200c})p(c_{200c}|M_{200c}) b(M_{200c},z) dM_{\rm{tot}} \nonumber\\
    A(M_1,z) &=& 1-I(M_1, k=0,z) \nonumber,
\end{eqnarray}
where $P_{\rm{lin}}(k,z)$ is the linear power spectrum of matter fluctuation, the linear halo bias $b(M_{200c},z)$ is modeled with the fitting function of \cite{Tinker2010}, $I(M,k,z)$ is the standard mean bias, and $M_1$ is still $10^{12}~h^{-1} M_\odot$ in $M_{200c}$. As the matter distribution is unbiased with respect to itself, the mean biases should approach 1 on large scales. We follow \cite{Mead2020} to add $A(M_1,z)$ in the mean bias calculation to guarantee this asymptotic behavior at $k=0$ given the fitting forms of \cite{Tinker2010} are normalized for $M=0$ and not $M=M_1$.

We can compute the suppression of the matter power spectrum $S(k,z)$ by taking a ratio of the following quantities:
\begin{equation}
\label{eq:suppresion}
    S(k,z) = \frac{P_{\rm{DMB}}(k,z)}{P_{\rm{NFW}}(k,z)} = \frac{P_{\rm{1H,DMB}}(k,z) + P_{\rm{2H}}(k,z)}{P_{\rm{1H,NFW}}(k,z) + P_{\rm{2H}}(k,z)}
\end{equation}
where $P_{\rm{1H,DMB}}(k,z)$ refers to the 1-halo term computed with the DMB model density from \ref{sec:theory}, and $P_{\rm{1H,NFW}}(k,z)$ is computed with a standard NFW profile. 
\subsection{Discussion}
\begin{figure}[htbp]
\centering
\includegraphics[width=\textwidth]{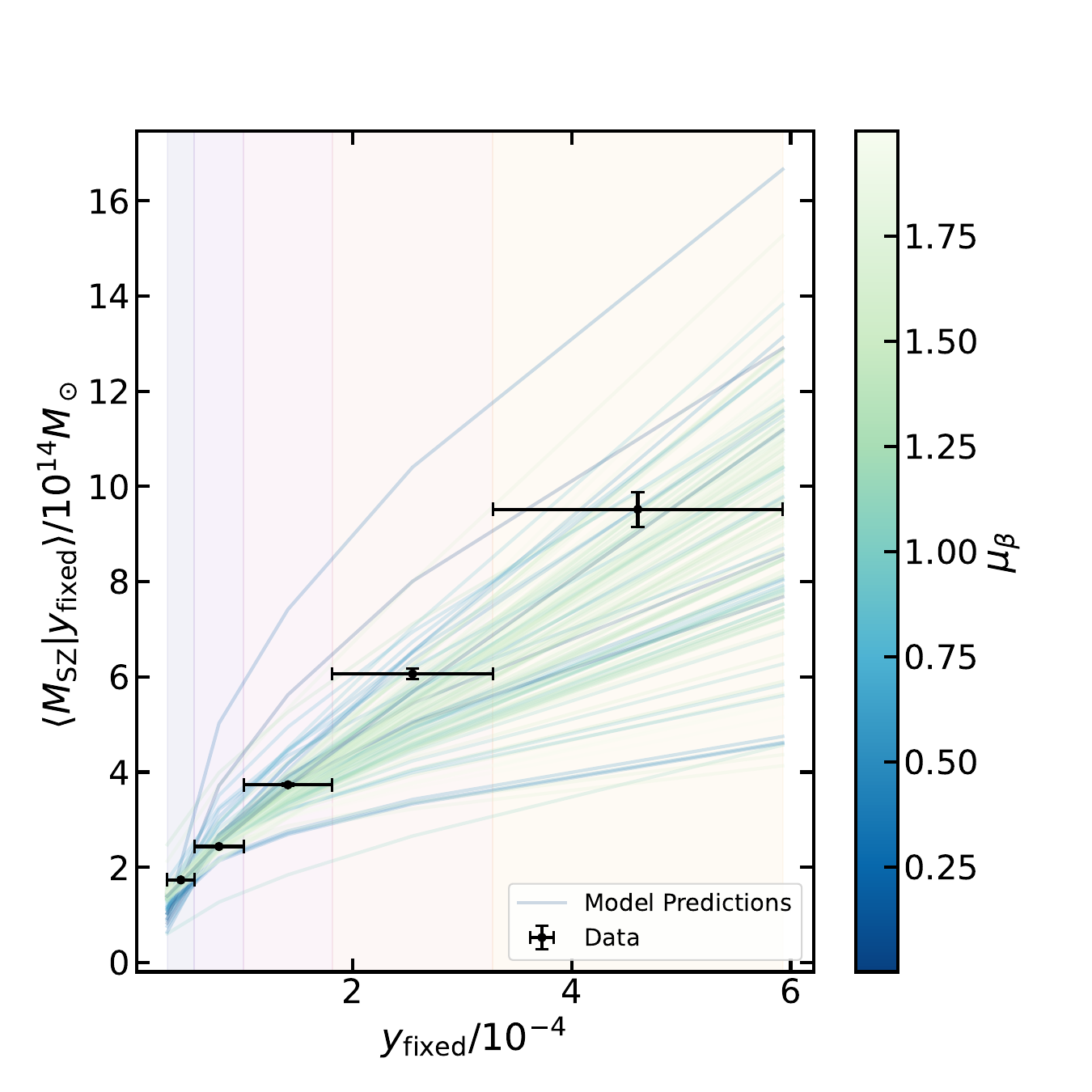}
\caption{Sample model fits to data, displayed with color according to the value of one of the better constrained parameters $\mu_\beta$, which is the mass dependence of the slope of the gas profile. The model fits are heavily weighted to the first three data points in $y_{\rm{fixed}}$ given that these three bins contain most of the clusters in the catalog, and have the tightest errors on the mean mass. The bin widths corresponding to the same bins in Figure \ref{fig:bins} are color coded in the background of the plot. \label{fig:v}}
\end{figure}

\begin{figure}[htbp]
\centering
\includegraphics[width=\textwidth]{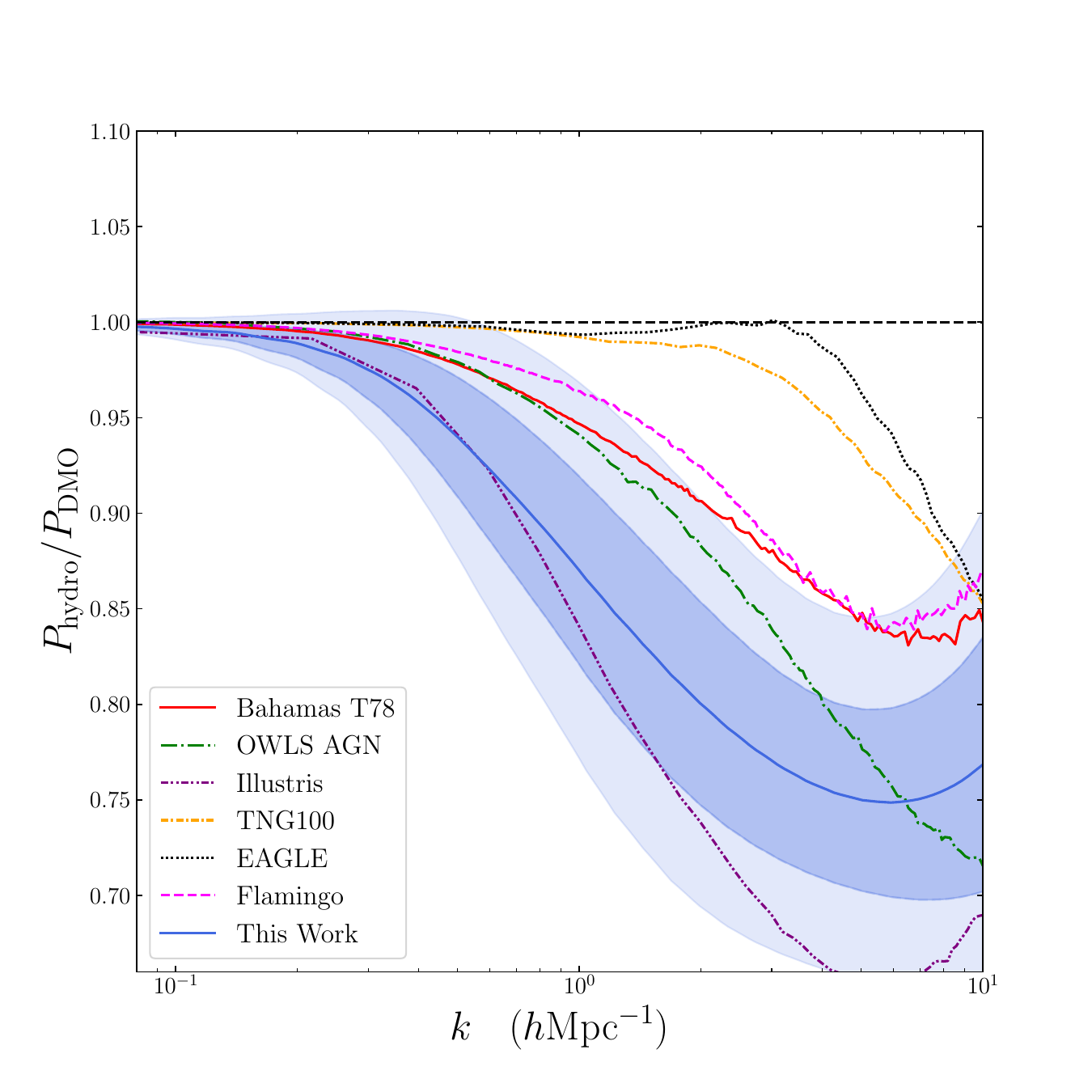}
\caption{Predicted suppression of the matter power spectrum assuming a Planck cosmology in comparison with suppressions measured from various hydrodynamical simulations. The darker shaded region presents our 68 \% confidence level, whereas the lighter shaded region shows our 95 \% confidence level.  \label{fig:iv}}
\end{figure}

We use Markov Chain Monte Carlo to find posterior distributions for our model parameters. We provide a cornerplot in Appendix \ref{sec:MCMC}. We down-sampled the chains and computed suppressions $S(k,z = 0)$ from the parameters using the halo model approach as detailed in Equation \ref{eq:suppresion}. Figure $\ref{fig:v}$ displays the corresponding downsampled model predictions compared to our datavector, whereas Figure $\ref{fig:iv}$ displays the distribution of the resulting suppression in the matter power spectrum in comparison to predictions from various hydrodynamical simulations. 

We find a preference for intermediate to strong feedback, which is consistent with the suppressions measured from several different cosmological scale hydrodynamical simulations such as Flamingo \cite{FlamingoSupp}, cosmoOWLS \cite{cosmoOWLS}, and Bahamas \cite{2017MNRAS.465.2936M}. However, we find discrepancies between our constraints and those measured in the EAGLE \cite{Eagle1, Eagle2} and TNG100 simulations \cite{2019ComAC...6....2N}, which both predict less extreme levels of feedback.

\begin{figure}[htbp]
\centering
\includegraphics[width=\textwidth]{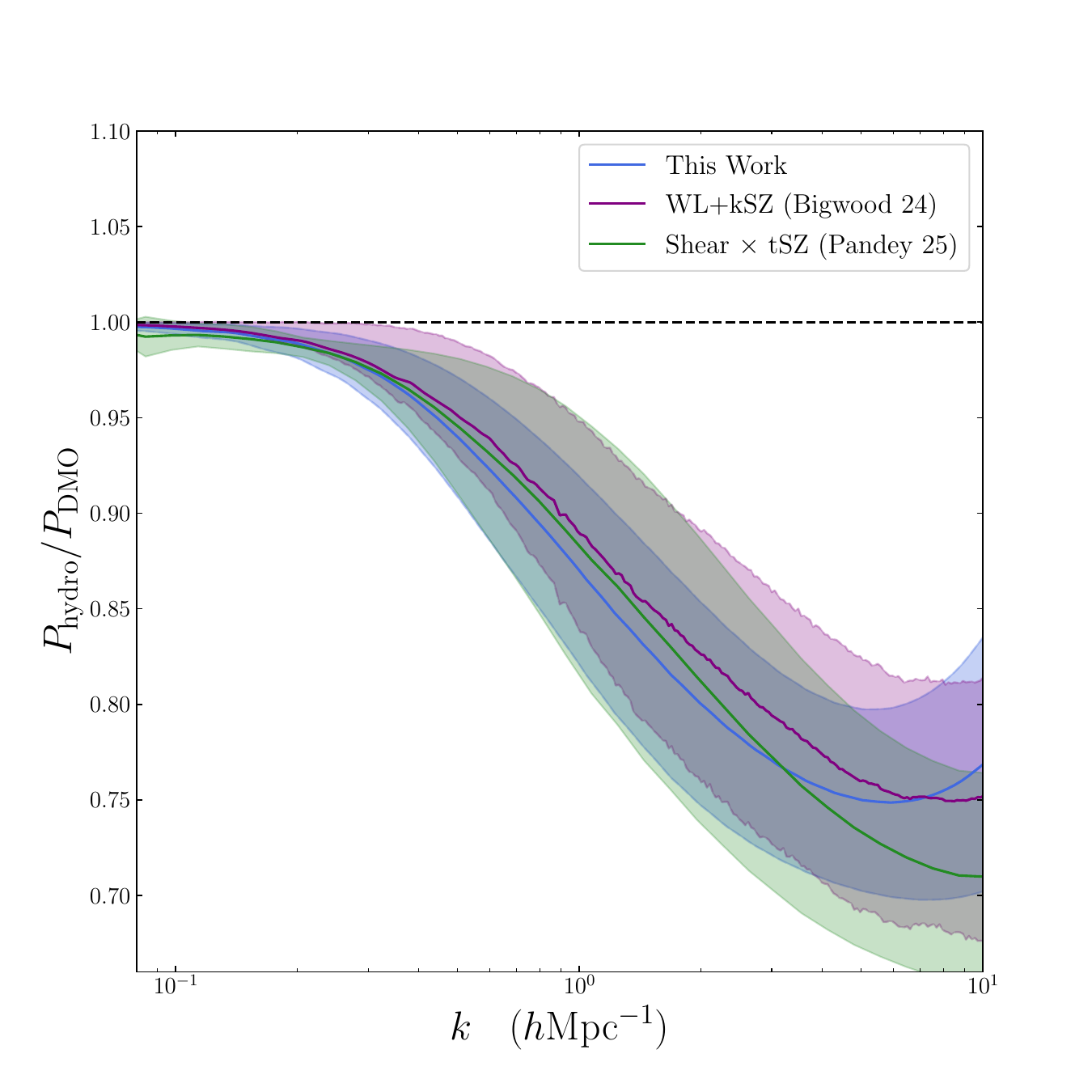}
\caption{Predicted suppression in comparison with constraints from DES weak lensing and ACT kSZ measurements \cite{DES:2024iny} and DES shear and ACT tSZ measurements \cite{shivam2025}. All suppressions were computed assuming a Planck cosmology \cite{Planck}.}  \label{fig:vi}
\end{figure}

In Figure \ref{fig:vi}, we plot our results at 68 \% confidence against the constrained suppression from \cite[hereafter B24]{DES:2024iny}, where the authors constrain $S(k)$  using kSZ measurement. We also compare to the suppression constrained in \cite[hereafter P25]{shivam2025}, which jointly models ACT DR6 tSZ signal and DES Y3 shear to constrain baryonic feedback. We find impressive consistency with both constraints on the suppression of the matter power spectrum. All three analyses constrain the gas parameters; the WL+kSZ study has somewhat tighter constraints on $\log_{10} M_{c0}$ and $\theta_{\rm{ej}}$, whereas our analysis has tighter constraints on $\mu_\beta$. We find this similarity particularly interesting given that tSZ clusters are more massive objects of mean $M_{\rm{200c}} \sim 3\times10^{14} M_{\odot}$, whereas the CMASS galaxies which \cite{DES:2024iny} used for velocity reconstructions have a mean halo mass of $M_{\rm{200c}} \sim 10^{13} M_{\odot}$. Furthermore, the kSZ effect directly probes the diffuse gas present in the exteriors of groups and clusters, whereas the tSZ effect is more prominent in the hot gas closer to the virial radius of the halo, and is jointly sensitive to both the temperature and the gas pressure. The median redshifts for our cluster sample and the kSZ CMASS sample are similar at $z \approx 0.55$. We also note that the model used by P25 is very similar to the model that we use in this analysis. Furthermore, given that both analyses are on ACT and DES data, the similarity in constrained suppressions is encouraging. 

Sustaining strong levels of feedback in such massive objects requires energetics and dynamics that we hope to explore in a future study but are beyond the scope of this paper. Our results further lend merit to a potential joint constraint combining all three observables (namely WL, tSZ, and kSZ), which would be extremely informative for constraining baryonic feedback across a wide range of massive objects that contribute to different scales in the cosmic shear signal. We also leave this for a future study.  

The most well constrained parameter from our analysis is $\mu_\beta$, which describes the mass dependence of the slope of the gas profile. We also have reasonably informative constraints on both $\theta_{\rm{ej}}$ and $\log_{10} M_{c0}$, which describe the pivot point and extent of the gas profile. We expect that this is the case because the tSZ signal extends well beyond the central galaxy component of the cluster, and therefore is uniquely informative about the gas parameters. While the constraints on other parameters remain prior-dominated, the constraints on the overall suppression are rather tight. This might seem surprising at first. Studies \cite{vanDaalenDiagram, zhou2025maplevelbaryonificationunifiedtreatment, Huang:2018wpy} have found that just a few parameters describing the gas content in halos are sufficient in predicting the suppression of matter power spectra due to baryonic physics, which hints that the parameter space relevant for constraining the suppression of the matter power spectrum is much smaller than the parameter space considered in our DMB halo model. In light of these results, one could consider varying only a few parameters, which will likely reduce the computational burden. We note, however, that in \citep{methodpaper}, we found that the parameters we considered are necessary to describe the gas profile in hydrodynamic simulations. We therefore adopt a conservative approach to vary all parameters when fitting to tSZ data. It is important to caveat that we use an informative prior on the stellar parameters $\eta_*$ and $\delta \eta$ from the results of \cite{Giri21}. 

\begin{figure}
    \centering
    \includegraphics[width=0.48\linewidth]{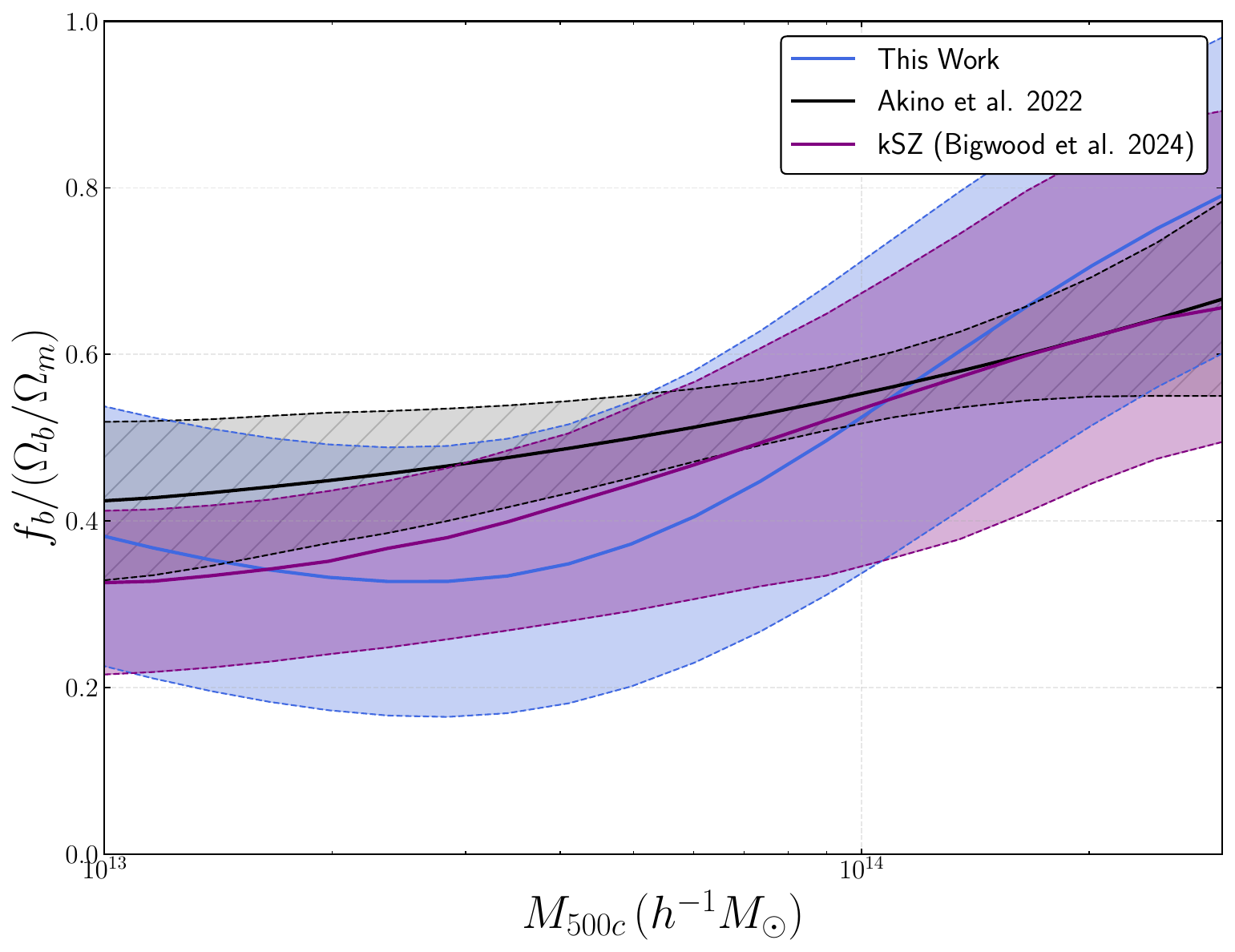}
    \includegraphics[width=0.48\linewidth]{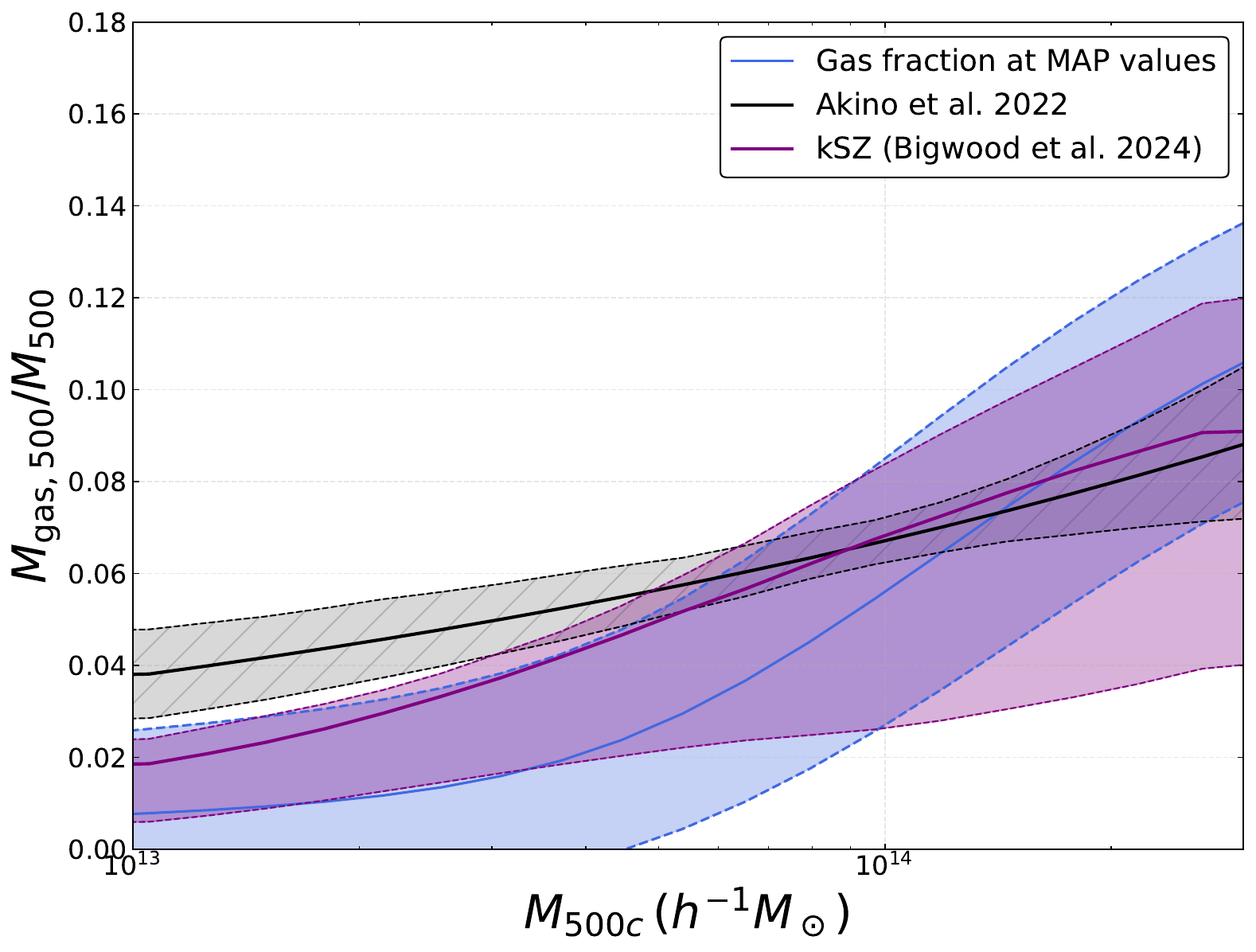}
    \caption{Left: the baryon fraction $f_b$ withing  $R_{\rm{500c}}$ as a ratio of the cosmic baryon fraction plotted against halo mass. Right: the fraction of gas mass within $R_{\rm{500c}}$ as a function of halo mass. In both plots, we plot the physical measurements from X-ray selected clusters \cite{2022PASJ...74..175A} in grey hatched, and the constraints from the kSZ study \cite{DES:2024iny} in purple. We present our results in blue, with the shaded region displaying the $1\sigma$ uncertainty. } 
    \label{fig:baryongasfracs}
\end{figure}

Furthermore, in Figure \ref{fig:baryongasfracs}, we plot the baryon and gas fractions ($f_{\rm{b}}/(\Omega_{\rm{b}}/\Omega_{\rm{m}})$ and $M_{\rm{gas}, 500}/M_{\rm{500c}}$ respectively) computed from samples drawn from the posterior as a function of halo mass $M_{\rm{500c}}$ at $z=0$. Note that $f_{\rm{b}}$ is specifically defined as the fraction of baryonic matter within $R_{\rm{500c}}$ of a given halo. Our constraints are in good agreement with observational measurements from X-ray selected clusters as measured by \cite{2022PASJ...74..175A}, with a suggestion of lower gas fractions below $M_{\rm{500c}} \sim 4 \times 10^{13} h^{-1} M_\odot$. We also plot the results from B24, which we remain consistent with. We note that our constraining power on both quantities is weaker than B24 as our model includes redshift evolution of the pivot mass scale with the parameter $\nu_z$, and that our data is most sensitive to the baryon content of clusters at a median redshift of $z \approx 0.55$. In extrapolating to $z=0$, we significantly decrease our confidence due to the relatively unconstrained redshift evolution of the model. More details are in Appendix \ref{sec:redshift}. We also find that despite the relatively wide error bars in these quantities at the low mass end, we constrain the suppression in the power spectrum at similar confidence as B24. This suggests that our ten-parameter DMB model has relevant information on the suppression of the power spectrum that is not quite captured in the baryon fractions in nearby halos. 

\begin{figure}
    \centering
    \includegraphics[width=1\linewidth]{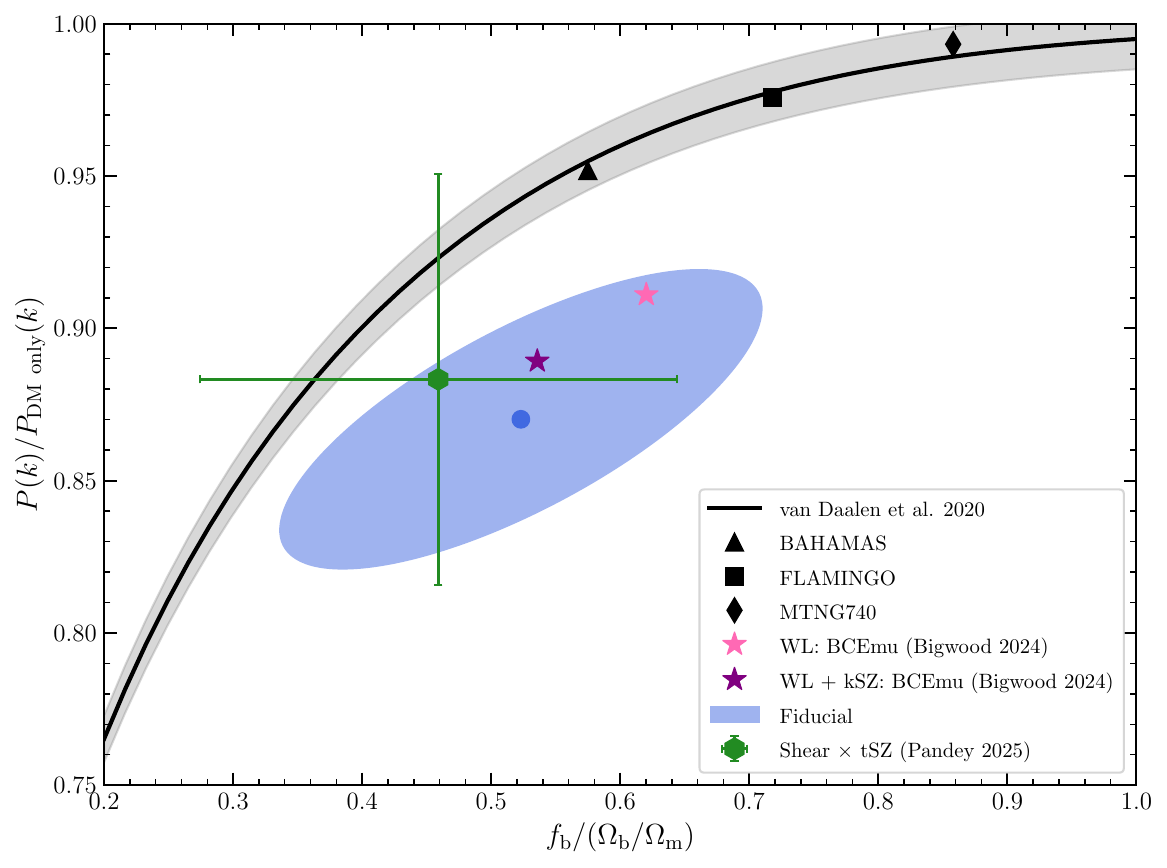}
    \caption{Our results plotted on the van Daalen diagram from \cite{vanDaalenDiagram}, which plots the baryon fraction in $10^{14}~h^{-1} M_\odot$ against the suppression in the power spectrum at $k = 1~h/\rm{Mpc}$. In pink and purple are the results from the kSZ measurement \cite{DES:2024iny}, in green is the DES-Y3 shear $\times$ ACT tSZ measurement \cite{shivam2025}, and in blue is our result, with a $1\sigma$ confidence region.}
    \label{fig:vanDaalen}
\end{figure}

Towards this end, we also plot our results on a van Daaelen diagram \cite{vanDaalenDiagram} in Figure \ref{fig:vanDaalen}, which plots the suppression of matter power spectrum at $k = 1~h^{-1} \rm{Mpc}$ against the baryon fraction $f_{\rm{b}}$ in halos of mass $M = 10^{14}~h^{-1} M_\odot$. Various hydrodynamic simulations fall on the tight empirical relation, as found by \cite{vanDaalenDiagram}, and as plotted in the figure in the black line. We plot our own results in blue, along with a $1\sigma$ confidence region. We find suggestions of a tension in the direction of stronger suppression at a given baryon fraction, which is also suggested by the results of B24.  

We constrain these results assuming a fixed Planck cosmology \cite{2016A&A...594A..13P}. We tested the cosmology dependence of the results by running chains at a DES cosmology, and found no noticeable difference in the predicted suppressions. We discuss this further in Appendix \ref{sec:CosmologyDependence}. 

\section{Conclusion}
\label{sec:conc}
Thermal SZ (tSZ) has been one of the primary signals for cluster finding in the millimeter sky. However, while previous studies have used it for ranking clusters, connecting the signal to the gas physics in galaxy clusters remains largely unexplored. Through various cross correlation studies, it has been demonstrated that the tSZ can provide powerful constraints for the gas physics around galaxies and clusters. In  \cite{Paper1}, we have shown that using the tSZ signal of galaxy clusters with weak-lensing mass calibrations will be a powerful tool to constrain baryonic physics in galaxy clusters, and therefore place a stringent constraint on the impact of baryonic suppression of the matter power spectrum which is crucial for cosmic shear anlalyses. 

In this paper, we employ the model presented in \cite{Paper1} to connect the tSZ measurement around ACT-DR5 clusters, whose mass is calibrated with DES-Y3 weak lensing, to the suppression of matter power spectra due to baryonic feedback. We are able to constrain this suppression to 5 \% at $k = 1 \, \rm{h/Mpc}$ and find a strong agreement with both the constraints from the kSZ measurement, as well as constraints from the recent shear $\times$ tSZ measurement. We further compare our constraints to predictions from various hydrodynamic simulations, and find that the suppression we constrain is in 3$\sigma$ tension with that predicted by IllustrisTNG. 

In the process of obtaining these constraints, we have developed an end-to-end pipeline that directly models the observables for tSZ clusters. We achieved this by building an emulator that predicts the observed signal as a function of the true signal under multiple variations of our model parameters. Although this approach was focused on the Atacama Cosmology Telescope, we hope to replicate it for other CMB experiments including the South Pole Telescope and the Simons Observatory. 

We have also demonstrated that the tSZ can place novel and informative constraints on DMB halo model parameters. In combination with other probes such as X-ray and kSZ, we can better constrain several different components of the halo model, and significantly reduce the prior volumes that would enter in a cosmic shear analysis. Of particular interest is calibrating the levels of feedback in objects of different mass ranges, where different mechanisms of feedback (see for example \cite{Quataert:2025vhp}) may become more prevalent. While our results show remarkable consistency between feedback calibrated at the group mass scale and the tSZ cluster scale, the mass dependence of feedback merits further investigation. 

While our constraint on the suppression of the matter power spectrum is competitive with other results, we expect further improvement in the near future. The ACT DR6 cluster catalog will double the cluster number counts and also enhance the certainty in the mass calibration through weak lensing. At the same time, the SPT-3G clusters \citep{2025arXiv250317271K} will further lower the mass range of the sample. Repeating our method on these datasets will further refine the constraints on the baryonic content of the halo, and the increased signal-to-noise ratio will enable more detailed studies, such as the redshift and mass evolution of the baryonic feedback. In the long term, we anticipate utilizing the constraints on baryonic feedback to unlock cosmological information from small-scale cosmic shear data. As shown in \cite{Paper1}, we anticipate that this analysis will significantly improve the cosmological constraints from cosmic shear data in DES and LSST-Y1.

\section{Data Availability}
We plan to make all the code and data products (including chains and computed suppressions) in this project available via a git repository after the paper is published. We will also include an example notebook that shows how various quantities can be computed. The repository will be hosted at \url{https://github.com/nihardalal/bemo-sz} 

\appendix
\section{Parameter Dependence}
\label{sec:ParameterDependence}
The dependence of the observable y-M relation and suppression on different model parameters is hard to decipher from the equations alone. In Figure \ref{fig:secA} below, we present the dependence of both quantities as they depend on the two best constrained parameters from our analysis, $\theta_{\rm{ej}}$ and $\mu_\beta$, which encode the extent of the gas distribution, and the mass dependence of the slope of the gas profile respectively.
\begin{figure}[htbp]
\centering
\includegraphics[width=\textwidth]{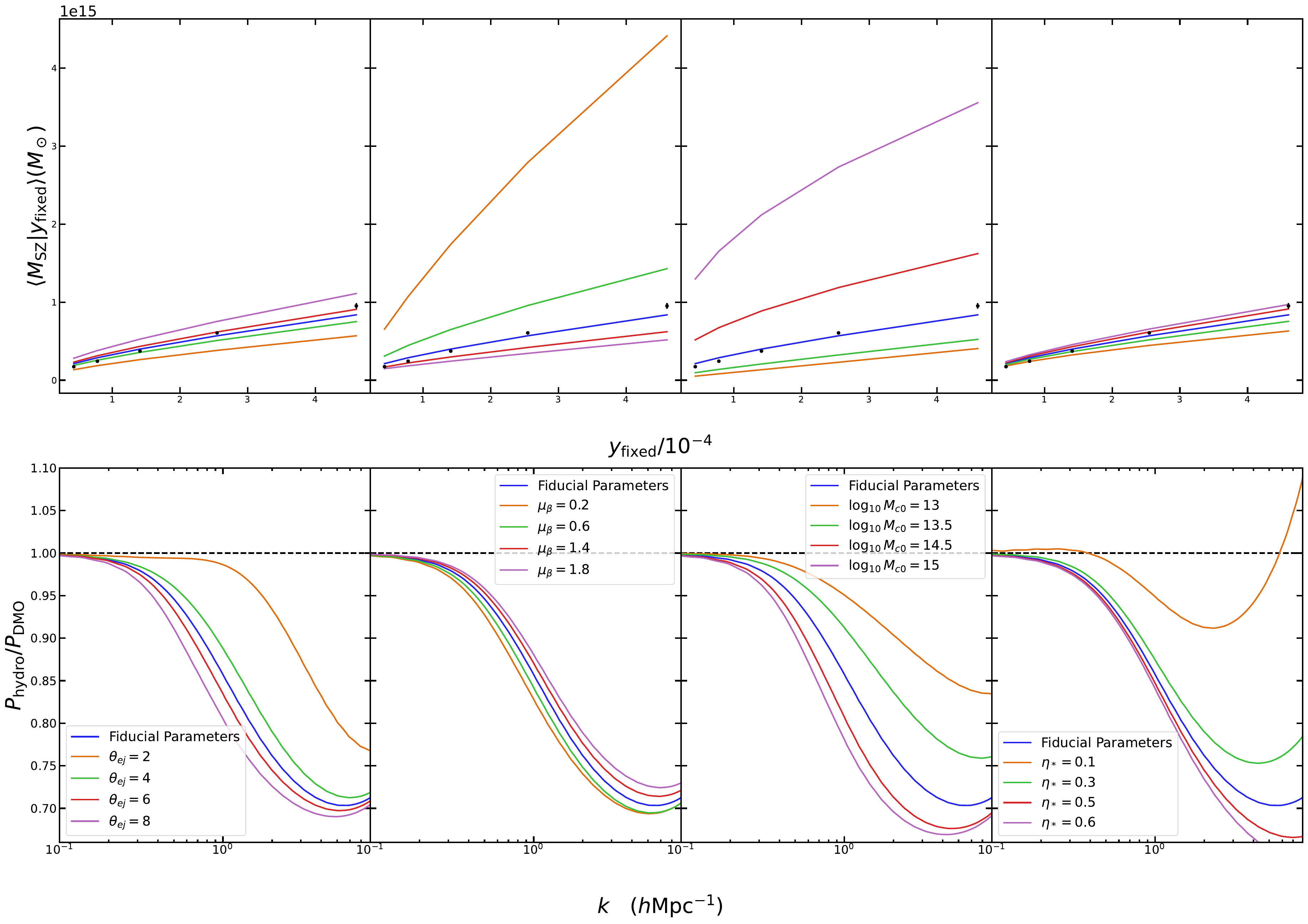}
\caption{ \label{fig:secA} Upper panels: the modeled observed y-M relation as it depends on $\theta_{\rm{ej}}$, $\mu_\beta$, $\log_{10} M_{c0}$, and $\eta_*$ (left to right). Lower panels: the dependence of the suppression of the matter power spectrum on $\theta_{\rm{ej}}$, $\mu_\beta$, $\log_{10} M_{c0}$, and $\eta_*$ (left to right). Fiducial values are the given by the midpoints of the priors in Table \ref{tab:Halomodel}.}
\end{figure}
\section{MCMC}
\label{sec:MCMC}
We used the \textsc{Zeus} \cite{karamanis2020ensemble, karamanis2021zeus} ensemble sampler to perform parameter inference on our models. The results are displayed below in Figure \ref{fig:ii}. 

It may seem somewhat counterintuitive that we are able to predict the suppression of the matter power spectrum rather well despite the fact that all parameters seem to be weakly constrained from the contour plot. In order to better understand the information constrained from the tSZ clusters, we perform a PCA analysis on the parameter covariance matrix of the posterior. 

We find one very well constrained parameter combination in the DMB model parameter space, with normalized eigenvalue of $0.16$, and corresponding eigenvector:
\begin{equation}
\begin{split}
\rm{PC1} = &-0.3009 \ln \theta_{\rm{ej}} +0.2602 \mu_\beta +0.1434 \ln \eta_* +0.2469 \ln \delta\eta +0.2806 \ln(\log_{10} M_{c0})\\ 
    & + 0.0274 \alpha_{\rm{nt}} -0.3114\nu_z -0.1874 \ln n_{\rm{nt}} +0.0435 \ln \gamma -0.6672\ln \delta +0.1642 \ln \sigma\\
    &+0.1528 \ln A_{\rm{mass}} +0.1766 \eta_{\rm{bias}}+0.1416\zeta_{\rm{bias}}
\end{split}    
\end{equation}
\begin{figure}[htbp]
\centering
\includegraphics[width=\textwidth]{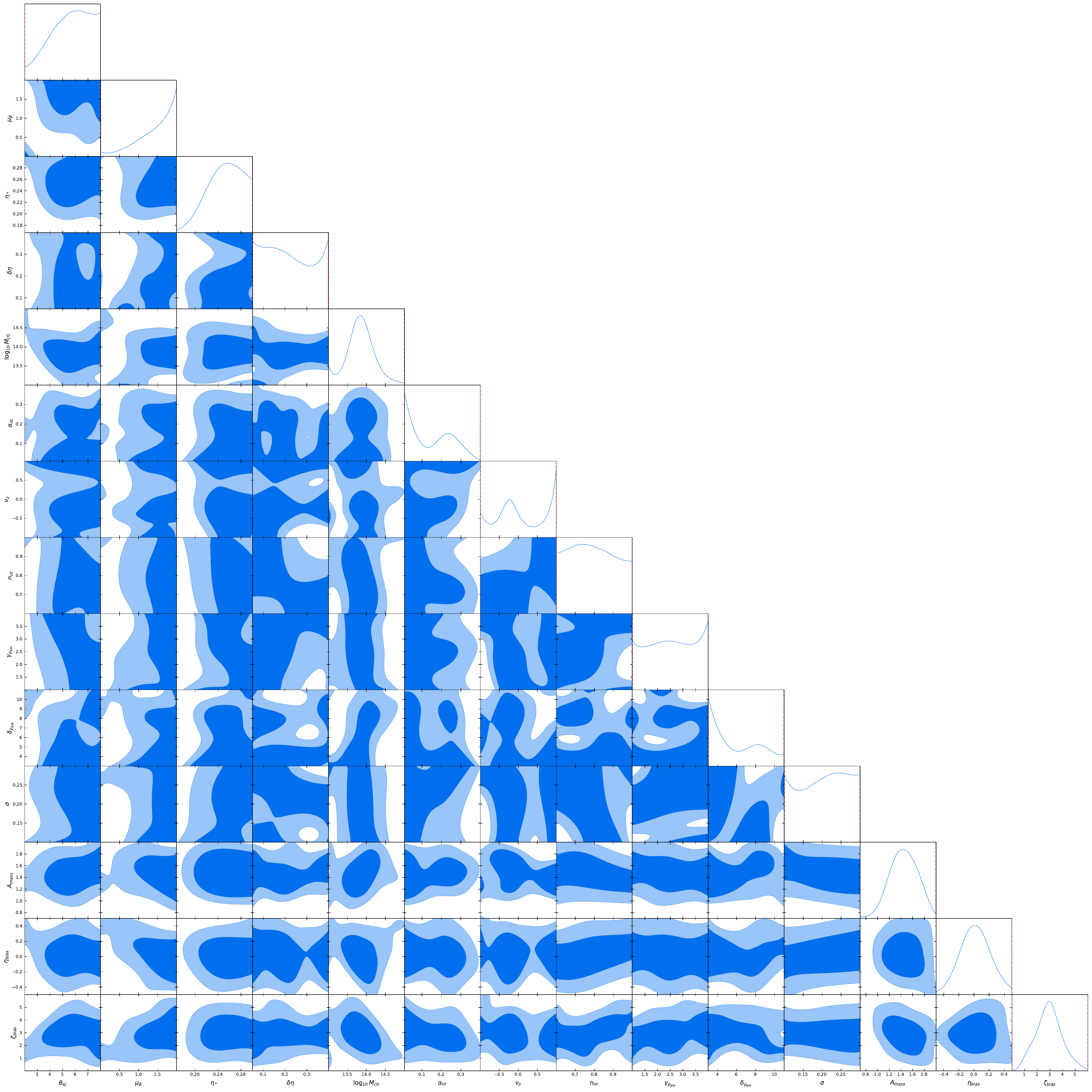}
\caption{Posterior distributions for model parameters with prior ranges described in Table \ref{tab:Halomodel}. We note that although the constraints on individual parameters may not seem tight, we can constrain the suppression rather well.  \label{fig:ii}}
\end{figure}
\section{Cosmology Dependence}
\label{sec:CosmologyDependence}
\begin{figure}[htbp]
\centering
\includegraphics[width=\textwidth]{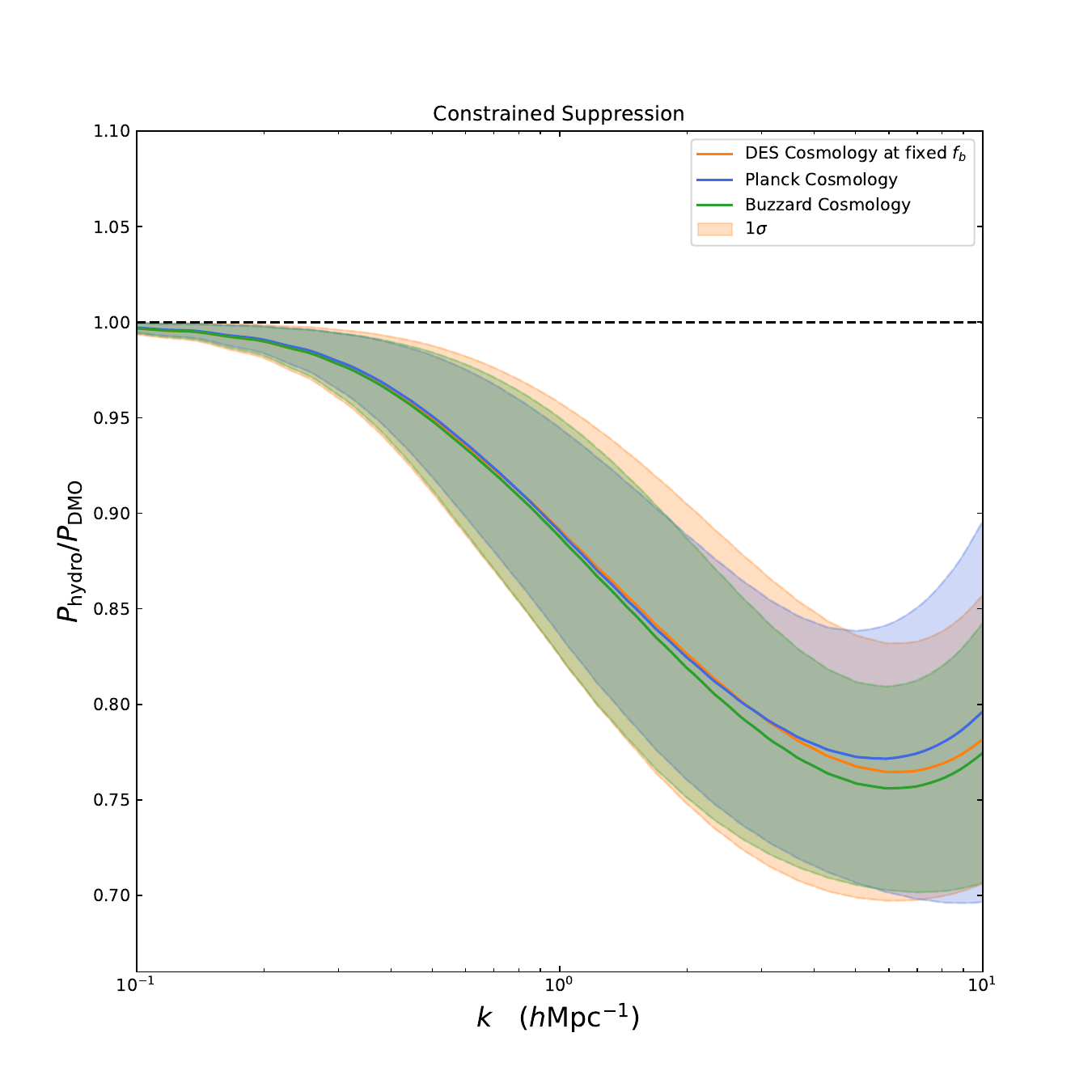}
\caption{Cosmology dependence of constrained suppressions, where the shaded regions display the 68 \% confidence levels. In blue is our fiducial Planck cosmology, in green we have a Buzzard cosmology, and in orange we have a best fit DES Y3 cosmology. All cosmologies have a fixed baryon fraction $f_{\rm{b}} = \Omega_{\rm{b}}/\Omega_{\rm{m}}$ \label{fig:viii}.}
\end{figure}

We briefly investigated the cosmology dependence of our model and constraints by running chains with different cosmologies, namely a Planck-like cosmology \cite{Planck}, a Buzzard-like cosmology, and a DES-like cosmology \cite{DESY3}. Varied parameters are presented in Table \ref{tab:cosmology}. We found that a majority of the dependence comes from the baryon fraction $f_{\rm{b}} = \Omega_{\rm{b}}/\Omega_{\rm{m}}$. Keeping $f_{\rm{b}}$ fixed, we find that the constrained suppressions at different cosmologies are consistent with one another, as shown in Figure \ref{fig:viii}. We did see a slight degradation in constraining power with the DES like cosmology, and we leave further testing of explicit cosmological dependence to future work.
\begin{table}[]
    \centering
    \begin{tabular}{c|c|c|c}
          & $\Omega_{\rm{m}}$ & $\sigma_8$ & $\Omega_{\rm{b}}$   \\
          \hline
         Planck & 0.3166  & 0.81 & 0.049\\
         Buzzard & 0.286 & 0.82 & 0.044\\
         DES & 0.339 & 0.776 & 0.052
    \end{tabular}
    \caption{Cosmological parameters varied to test the cosmology dependence of our results}
    \label{tab:cosmology}
\end{table}

\section{Redshift Dependence}
\label{sec:redshift}
\begin{figure}
    \centering
    \includegraphics[width=\linewidth]{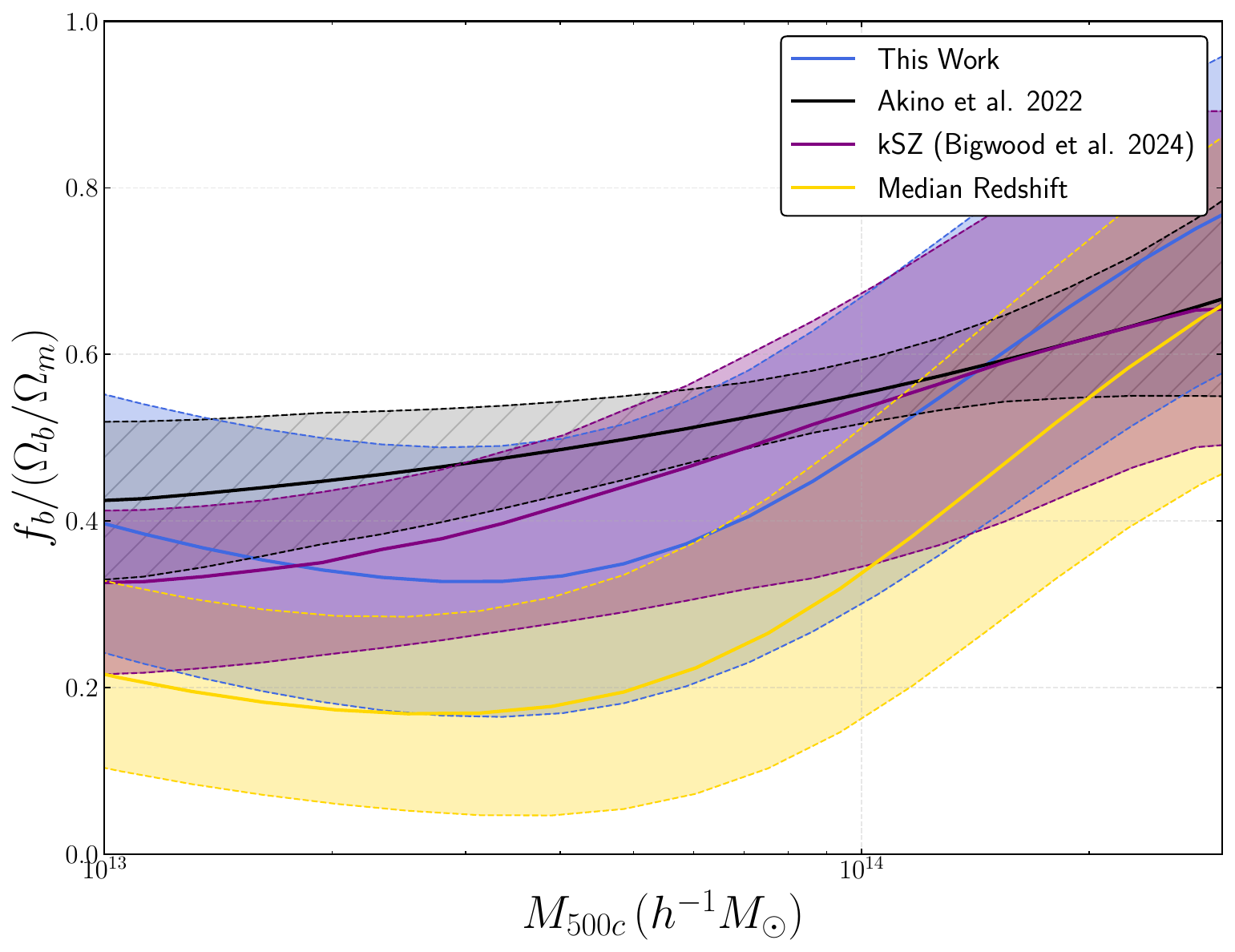}
    \caption{Baryon fractions, as plotted in Figure \ref{fig:baryongasfracs}, but with the addition of the baryon fraction computed at the median redshift of our catalog at $z\approx 0.55$ plotted in yellow. We note that the error is much lower compared to the error in the extrapolated quantity computed at $z=0$ in blue, because of the lack of constraints on $\nu_z$. These error levels are comparable to the errors of B24, plotted in purple.}
    \label{fig:redshiftevol}
\end{figure}

\begin{figure}
    \centering
    \includegraphics[width=\linewidth]{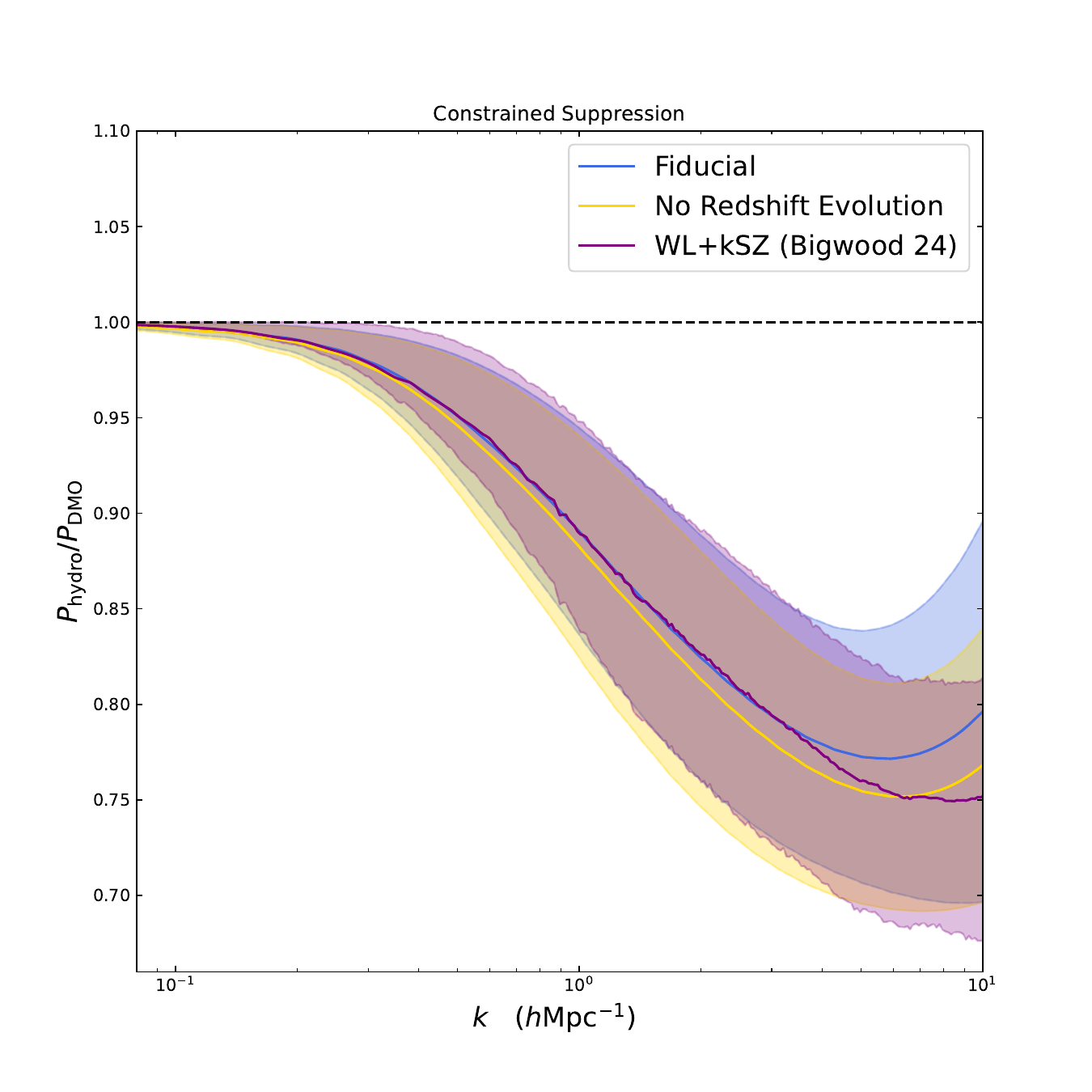}
    \caption{We plot the constrained suppression assuming no redshift evolution in the model in yellow, i.e. setting $\nu_z = 0$. We see that there is little effect of the redshift evolution on the level of suppression, and slightly better constraining power compared to the fiducial model, plotted in blue, and slightly better constraining power than B24, plotted in purple.}
    \label{fig:suppredshift}
\end{figure}

We note that the redshift evolution of our model, dictated by the parameter $\nu_z$ reduces certainty in the prediction of baryon and gas fractions in nearby halos, as shown in Figure \ref{fig:redshiftevol}, where we first check the baryon fraction at the median redshift of our cluster sample, at $z\approx 0.55$. We see that the baryon fraction is better constrained at this redshift, suggesting that there may be a loss in constraining power due to $\nu_z$. We therefore choose to recompute the suppression with $\nu_z = 0$. Results are shown in Figure \ref{fig:suppredshift}. We find that the suppression itself isn't significantly affected by this choice. We leave the redshift dependence of the model and the van Daalen relation to future studies, as they certainly merit further investigation beyond the scope of this paper. 

\acknowledgments

During the preparation of this work, N.D. and C.H. were supported by the David \& Lucile Packard Foundation; and by the National Aeronautics and Space Administration (Grant 22- ROMAN110011). CHT was supported by the Eric and Wendy Schmidt AI in Science Postdoctoral Fellowship, a Schmidt Futures program. CHT would like to acknowledge the Aspen Center for Physics, which is supported by National Science Foundation grant PHY-2210452. Part of the work was performed at the "Groups and Clusters of Galaxies at the Crossroad between Astrophysics and Cosmology" conference at the Aspen Center. We would like to thank the DES and ACT collaboration for their valuable input to this project. We'd also like to thank Leah Bigwood and David H. Weinberg for valuable discussions and assistance with this project. 

\bibliographystyle{JHEP}
\bibliography{biblio.bib}

\end{document}